\documentclass[12pt,preprint]{aastex}
\usepackage{emulateapj5}

\newcommand{\cf}{{\ifmmode{C_f}\else{$C_{f}$}\fi}}
\newcommand{\zem}{{\ifmmode{z_{em}}\else{$z_{em}$}\fi}}
\newcommand{\zabs}{{\ifmmode{z_{abs}}\else{$z_{abs}$}\fi}}
\newcommand{\kms}{{\ifmmode{{\rm km~s}^{-1}}\else{km~s$^{-1}$}\fi}}
\newcommand{\cmm}{cm$^{-2}$}

\newcommand{\voff}{{\ifmmode{v_{shift}}\else{$v_{shift}$}\fi}}
\newcommand{\ergs}{{\ifmmode{{\rm erg~s}^{-1}}\else{erg~s$^{-1}$}\fi}}
\newcommand{\rl}{{\cal R}}
\def\lsim{\lower0.3em\hbox{$\,\buildrel <\over\sim\,$}}
\def\gsim{\lower0.3em\hbox{$\,\buildrel >\over\sim\,$}}

\newcounter{species} 
\def\ion#1#2{\setcounter{species}{#2}#1$\;${\scriptsize\Roman{species}}\relax}

\newcommand{\hi}{{\rm H}~{\sc i}}
\newcommand{\lya}{Ly$\alpha$}

\slugcomment{\today}
\slugcomment{A complete version can be obtained by http://www.astro.psu.edu/users/misawa/pub/Paper/census.ps.gz}
\shorttitle{Census of Intrinsic Absorption Lines}
\shortauthors{Misawa et al.}

\begin{document}

\title{A Census of Intrinsic Narrow Absorption Lines in the Spectra of
Quasars at $z=2$--4\altaffilmark{1}}

\footnotetext[1]{The data presented here were obtained at the
  W.M. Keck Observatory, which is operated as a scientific partnership
  among the California Institute of Technology, the University of
  California and the National Aeronautics and Space
  Administration. The Observatory was made possible by the generous
  financial support of the W. M. Keck Foundation.}

\author{Toru Misawa\altaffilmark{2}, 
        Jane C. Charlton\altaffilmark{2}, 
        Michael Eracleous\altaffilmark{2,3}, 
        Rajib Ganguly\altaffilmark{4}, 
        David Tytler\altaffilmark{5,6},
        David Kirkman\altaffilmark{5,6},
        Nao Suzuki\altaffilmark{5,6}, and
        Dan Lubin\altaffilmark{5,6}
}

\altaffiltext{2}{Department of Astronomy and Astrophysics, The
  Pennsylvania State University, 525 Davey Lab, University Park, PA
  16802}

\altaffiltext{3}{Center for Gravitational Wave Physics,
  The Pennsylvania State University, University Park, PA 16802}

\altaffiltext{4}{Department of Physics and Astronomy, 1000 East
  University Ave, University of Wyoming (Dept 3905), Laramie, WY,
  82071}

\altaffiltext{5}{Center for Astrophysics and Space Sciences,
  University of California San Diego, MS 0424, La Jolla, CA
  92093-0424}

\altaffiltext{6}{Visiting Astronomer, W. M. Keck Observatory, which is
  a joint facility of the University of California, the California
  Institute of Technology, and NASA}

\begin{abstract}

We use Keck/HIRES spectra of 37 optically bright quasars at $z=2$--4
to study narrow absorption lines that are intrinsic to the quasars
(intrinsic NALs, produced in gas that is physically associated with
the quasar central engine). We identify 150 NAL systems, that contain
124 \ion{C}{4}, 12 \ion{N}{5}, and 50 \ion{Si}{4} doublets, of which
18 are associated systems (within 5,000 \kms\ of the quasar
redshift). We use partial coverage analysis to separate intrinsic NALs
from NALs produced in cosmologically intervening structures.  We find
39 candidate intrinsic systems, (28 reliable determinations and 11
that are possibly intrinsic). We estimate that 10--17\% of \ion{C}{4}
systems at blueshifts of 5,000--70,000~\kms\ relative to quasars are
intrinsic.  At least 32\% of quasars contain one or more intrinsic
\ion{C}{4} NALs. Considering \ion{N}{5} and \ion{Si}{4} doublets
showing partial coverage as well, at least 50\% of quasars host
intrinsic NALs. This result constrains the solid angle subtended by
the absorbers to the background source(s).  We identify two families
of intrinsic NAL systems, those with strong \ion{N}{5} absorption, and
those with negligible absorption in \ion{N}{5}, but with partial
coverage in the \ion{C}{4} doublet.  We discuss the idea that these
two families represent different regions or conditions in accretion
disk winds.  Of the 26 intrinsic \ion{C}{4} NAL systems, 13 have
detectable low-ionization absorption lines at similar velocities,
suggesting that these are two-phase structures in the wind rather than
absorbers in the host galaxy. We also compare possible models for
quasar outflows, including radiatively accelerated disk-driven winds,
magnetocentrifugally accelerated winds, and pressure-driven winds, and
we discuss ways of distinguishing between these models
observationally.

\end{abstract}

\keywords{accretion -- quasars: absorption lines --- quasars: general
  --- galaxies: active}

\section{Introduction\label{sec:intro}}

Quasars and active galactic nuclei (AGNs) are thought to be powered by
accretion of matter onto a supermassive black hole. The accretion flow
onto the black hole is thought to proceed via an equatorial accretion
disk, which provides a mechanism for removing the angular momentum of
the infalling matter. Outflows (winds) from such disks appear to be an
inseparable part of this process. More specifically, hydromagnetic
winds (e.g., Blandford \& Payne 1982; Emmering, Blandford \& Shlosman
1992; Konigl \& Kartje 1994; Everett 2005) provide a potential
mechanism for extracting angular momentum from the accreting material,
thus allowing accretion to proceed. Accretion disk winds are likely to
be responsible for the broad absorption lines (BALs) observed in the
spectra of a fraction of quasars (e.g., Murray et al. 1995; Arav, Li,
\& Begelman 1994, Proga, Stone, \& Kallman 2000) and may even be the
source of the broad emission lines that are the hallmark of all
quasars and AGNs (e.g., Chiang \& Murray 1996; Murray \& Chiang 1997).
Understanding the quasar outflow mechanisms by testing and refining
theoretical models is therefore an essential part of our quest to
understand the central engines of AGNs and quasars.

Quasar outflows also have quite important consequences for cosmology
and galaxy formation and evolution because they provide energy and
momentum feedback to the interstellar medium (ISM) of the host galaxy
and to the intergalactic medium (IGM). Simulations of galaxy evolution
through mergers (Springel, Di Matteo \& Hernquist 2005) show that AGN
feedback heats the ISM and inhibits star formation. Thus the colors of
the resulting galaxies evolve very quickly to the red, in agreement
with the observed color distribution of nearby galaxies. In
semi-analytic models of galaxy assembly (Granato et al. 2004;
Scannapieco \& Oh 2004) AGN feedback expells dense gas from the
centers of the host galaxies, heating up the IGM (which it also
enriches with metals). Because of its low density, the IGM does not
cool efficiently and as a result it cannot fall back onto the galaxy
and fuel star formation. Thus feedback brings about an early
termination of the assembly of the host galaxy. In this context it is
quite important to know the fraction of quasars driving energetic
outflows, as well as the energy content of the outflow. The latter
quantity can be determined from the velocity, column density, and
global covering factor of the outflowing gas. All of these parameters
are intimately connected to dynamical models of accretion disk winds
and constraining them observationally can hardly be overemphasized.
Moreover, the outflowing material may provide a mechanism for
enriching the IGM with metals. This idea is made more attractive by
recent results that suggest high metallicities in quasar outflows
(e.g., Hamann et al. 1997; Gabel et al. 2006; Petitjean et al. 1994;
Tripp et al. 1996; D'Odorico et al. 2004). This question is closely
connected to the issues above since the outflow rates and
metallicities of quasar winds are at the heart of the matter.

In the models cited above, the outflow is accelerated by either
magnetocentrifugal forces, or by radiation pressure in lines and
continuum, or by a combination of the two processes. The flow
originates deep inside the potential well of the black hole and can
reach terminal speeds of order $10^4~\kms$ or higher. The geometry of
the flow differs between a purely radiation-driven and a purely
magnetocentrifugal wind. In the former case, (and especially at high
luminosities) the flow is largely confined to low latitudes and
effectively ``hugs'' the accretion disk. In addition, the numerical
simulations of Proga et al. (2000) show that the region above the
fast, low-latitude stream develops transient filaments and streams
which are denser than the ambient medium.  In contrast, in the latter
case the flow is nearly cylindrical (or U-shaped) and appears somewhat
collimated at large radii from the black hole (see, for example, the
illustrations in Blandford \& Payne 1982 and Konigl \& Kartje
1994). However, the density of the flow drops sharply once the gas has
traveled a distance of a few launch radii, with the consequence that
the dense part of the flow is concentrated near the equatorial plane
(e.g., Everett 2005).  A hybrid model, in which radiation pressure and
magnetocentrifugal forces are combined, also leads to a similar
geometry (Everett 2005). One may expect by analogy with radiation
pressure-driven winds that the higher-latitude, lower-density regions
of these winds would also contain transient, dense filaments or
streams. Thus the {\it apparent} geometry and kinematics of the flows
resulting from these two acceleration mechanisms are fairly similar.

A different family of outflow models includes pressure-driven winds
(e.g., Balsara \& Krolik 1993; Krolik \& Kriss 1995, 2001; Chelouche
\& Netzer 2005). These are made up of gas that is photo-evaporated
from the cool, dense torus that is invoked in AGN unification
schemes. Thus, the flow originates not very deep inside the black hole
potential well and the outflow speed is of order $10^3~\kms$.  The
outflow is expected to have a multi-temperature and multi-density
structure and indeed it would be clumpy. The primary difference
between this type of model and the accretion-disk wind models
described above is the terminal speed of the flow.

Blueshifted (or ``P-Cygni'') absorption lines provide a direct probe
of quasar accretion disk winds. BALs (widths $> 2,000$~\kms, by
definition) have been the traditional means of studying such winds
because they can be readily associated with them. This association is
motivated by their large widths (often corresponding to velocity
spreads of 30,000~\kms) and their smooth profiles.  However, intrinsic
narrow absorption lines (intrinsic NALs; widths up to 500~\kms; see
review by Hamman \& Sabra 2004) that are physically related to the
quasars are an alternative and very useful means of studying such
outflows. Intrinsic (most often blueshifted) NALs are detected in a
significant fraction (25--50\%) of quasar spectra. These correspond to
UV resonance transitions in highly ionized ions, such as \ion{C}{4}
and \ion{N}{5} (e.g., Foltz et al 1986, hereafter F86; Sargent,
Steidel, \& Boksenberg 1989; Anderson et al. 1987; Young, Sargent, \&
Boksenberg 1982; Steidel \& Sargent 1991; Ganguly et al. 2001,
hereafter G01; Vestergaard 2003). Quasar spectra also host an
intermediate class of absorption lines, mini-BALs (widths between 500
and 2,000~\kms, see Hamann \& Sabra 2004), whose widths and smooth
profiles also suggest an origin in outflows. Such absorption lines are
also found in Seyfert 1 galaxies (e.g., Crenshaw, Maran, \& Mushotzky
1998; Crenshaw et al. 2004; Scott et al. 2004).  Studies of intrinsic
NALs and mini-BALs are more promising than studies of BALs for two
important reasons: first these lines do not suffer from self-blending,
and second, they are found in a wider variety of active galactic
nuclei, and with greater ubiquity. In the context of the models
described above, intrinsic NAL may probe the dense, cool filaments
embedded in a hotter outflow.  Thus they may give us access to a
different portion, a different phase, or a different line of sight
through the outflow than the BALs.

In spite of their potential importance, NALs have not received as much
attention as BALs. Thus, we know surprisingly little about the
relation between intrinsic NALs, mini-BALs, and BALs, and their
connection, if any to the properties of the broad emission-lines.
This is because it is difficult to distinguish intrinsic NALs from
NALs that are not physically related to the quasars (intervening
NALs), produced in intervening galaxies, intergalactic clouds, Milky
Way gas or gas in the host galaxies of the quasars. It has been
traditionally thought that NALs that fall within 5,000~\kms\ of quasar
emission redshifts (termed {\it associated} absorption lines or AALs)
are physically associated with the quasars, because their frequency
per unit velocity increases with decreasing velocity offset from the
quasar (e.g., Weymann et al. 1979). AALs are common in all types of
quasars, although the strongest ones appear preferentially in
radio-loud quasar spectra (F86; Anderson et al. 1987). The statistical
analysis of Richards et al. (1999) and Richards (2001) suggested that
a fraction of 36\% of the NALs with blueshifts from 5,000 to
65,000~\kms\ from a quasar may also be physically associated with it.

Over the past decade, with the advent of high-dispersion spectroscopy
of faint objects, it has become possible to identify individual
intrinsic NALs, based primarily on one or both of the following
indicators: (a) the dilution of absorption troughs by unocculted light
(e.g., Hamann et al. 1997a; Barlow \& Sargent 1997; Ganguly et
al. 1999, hereafter G99; Ganguly et al. 2003; Misawa et al. 2003), and
(b) time variability of line profiles (e.g., depth, equivalent width,
and centroid), within a year in the absorber's rest frame (e.g.,
Barlow \& Sargent 1997; Wampler, Chugai, \& Petitjean 1995; Hamann,
Barlow, \& Junkkarinen 1997b; Wise et al. 2004; Narayanan et al. 2004;
Misawa et al. 2005, hereafter M05).

In this paper, we use the former of the above techniques to identify
intrinsic \ion{C}{4}, \ion{Si}{4}, and \ion{N}{5} NALs in the
high-resolution spectra of 37 quasars at $\zem\ = 2$--4. These quasars
were originally selected without regard to NAL properties, though BAL
quasars were avoided, and there was a preference for optically bright
quasars.  These data allow us to construct a large, and relatively
unbiased, sample of intrinsic NALs.  Using this sample, we investigate
the NAL properties statistically and we compare them with the
properties of the quasars that host them. Since our spectra typically
cover several transitions of the same system, we are also able to
probe the ionization state of the absorber. This large sample
represents a major advance over most previous efforts, which either
dealt with small samples of intrinsic NALs or employed low-resolution
spectra in statistical studies.

In \S\ref{sec:obs} we describe the properties of the quasar sample and
briefly summarize the observations.  In \S\ref{sec:sample} we describe
our methodology for identifying intrinsic NALs and evaluating the
reliability of this determination. Our results are presented in
\S\ref{sec:results}, and their implications in the context of models
for quasar outflows are discussed in \S\ref{sec:discussion}. Our
conclusions are summarized in \S\ref{sec:summary}. In this paper, we
use a cosmology with $H_{0}$=75 \kms Mpc$^{-1}$, $\Omega_{m}$=0.3, and
$\Omega_{\Lambda}$=0.7. In the Appendix we present all the detailed
information on the NALs we have detected.  Included in this Appendix
are (a) a table of NAL properties derived from fits to their profiles,
(b) diagnostic plots of UV resonance doublets, which form the basis of
our intrinsic NAL identification method, and (c) comparison plots of
profiles of different transitions from the same intrinsic NAL system,
(d) detailed notes on individual intrinsic NAL systems.

\section{Quasar Sample and Observations\label{sec:obs}}

The quasars in our sample were originally selected and observed in a
survey aimed at measuring the deuterium-to-hydrogen abundance ratio
(D/H) in the \lya\ forest. The typical value of D/H is so small,
2--$4\times 10^{-5}$ (O'Meara et al. 2001 and references therein),
that we can detect only \ion{D}{1} lines corresponding to \hi\ lines
with large column densities, $\log$($N_{\rm H\;I}$/\cmm) $\geq 16.5$.
Therefore the survey included 40 quasars, in which either damped \lya\
(DLA) systems or strong Lyman limit systems (LLSs) were detected. The
observations were carried out with Keck/HIRES through a 1{\farcs}14
slit resulting in a velocity resolution of $\sim 8$~\kms\ (FWHM). The
spectra were extracted by the automated program, MAKEE, written by Tom
Barlow.  In this paper we use the spectra of 37 of these quasars,
listed in Table~\ref{tab:quasars}, which cover the rest-frame
wavelength range between the \lya\ and \ion{C}{4} lines.  This target
selection method does not directly bias our sample with respect to the
properties of any {\it intrinsic} absorption-line systems that these
spectra may contain.  However, an indirect bias could result since the
sample contains only quasars bright enough to allow high $S/N$, high
resolution observations.  Although, optical brightness does not appear
to be the most significant factor in determining whether a quasar
hosts a NAL, it is likely to have some effect (G01).  Also, there is a
somewhat enhanced probability of finding associated NALs in quasars
that host BALs (G01), thus the avoidance of BAL quasars in our sample
may bias against intrinsic NALs.  We will discuss these possible
biases, and comparisons to other samples selected by different
criteria, in \S~\ref{sec:velew}.

Quasar emission redshifts were obtained from a variety of sources in
the literature and they are based primarily on measurements of the
peaks of their strong, broad UV emission lines, namely \lya,
\ion{Si}{4} and \ion{C}{4}. It is well known that the redshift
determined from these particular lines is systematically different
from the redshift of the low-ionization lines (i.e., \ion{Mg}{2} and
the Balmer lines) and the systemic redshift, as given by the narrow,
forbidden lines (see, for example, Corbin 1990; Tytler \& Fan 1992;
Brotherton et al. 1994; Sulentic et al. 1995; Marziani et
al. 1996). Even though redshift differences can reach $\sim
4,000~\kms$, in most cases they are within $\pm 1,000~\kms$. In
radio-quiet quasars, there is a systematic tendency for the \ion{C}{4}
line to be {\it blueshifted} relative to the systemic redshift. Most
relevant to this work are the results of Tytler \& Fan (1992) who
study the redshift discrepancies of the UV lines in quasars of
comparable redshifts and luminosities to those of our sample. They
find a mean blueshift of the UV lines relative to the systemic
redshift of 260~\kms\ and that 90\% of the blueshifts are between 0
and 650~\kms\ (see their Figure~13).

We also list in Table~\ref{tab:quasars} a measure of the
radio-loudness of the quasars in our sample, the ratio of the flux
densities at 5~GHz and 4400~\AA, i.e., $\rl = f_{\nu}(5\;{\rm
GHz})/f_{\nu}(4400\;{\rm\AA})$ (following Kellermann et al. 1989,
1994). We adopt $\rl \geq 23$ as the criterion for radio-loudness
(instead of the $\rl \geq 10$ adopted by Kellermann et al.)  in order
to separate cleanly the low-luminosity radio sources in our sample
from the high-luminosity ones, which have $\rl > 252$.  We derived
$f_{\nu}(4400\;{\rm\AA})$ from the $V$ or $R$ magnitude using the
following equations (Schmidt \& Green 1983; Oke \& Schild 1970),
\begin{eqnarray}
 m_V = & -2.5\log f_{\nu}(5500{\rm \;\AA}) - 48.60 & \label{eqn:vmag} \\
 m_R = & -2.5\log f_{\nu}(6600{\rm \;\AA}) - 48.82 & \label{eqn:rmag}
\end{eqnarray}
and assuming an optical spectral index of $\alpha_o = 0.44$ (where
$f_{\nu} \propto \nu^{-\alpha}$; see Vanden~Berk et al. 2001). For
Q0241$-$0146 (\zem\ = 4.040) and Q1055$+$4611 (\zem\ = 4.118), the
observed optical fluxes obtained from the $V$ magnitude are
underestimated because of contamination by the \lya\ forest at \zabs
$\sim$ 3.5. Therefore, we boosted these fluxes by dividing by the
normalized transmission in the \lya\ forest at redshift $z$,
\begin{equation}
  T = \exp\left[-0.0037\; (1+z)^{3.46}\right]
\end{equation}
(Press, Rybicki, \& Schneider 1993). The value of $f_{\nu}(5\;{\rm
GHz})$ was obtained from measurements of the flux density at various
radio frequencies, assuming a radio spectral index of $\alpha_r =
0.7$.  Thus, our 37 quasars are separated into 12 radio-loud quasars
and 25 radio-quiet quasars. Unfortunately, the number of radio-loud
quasars that cover the necessary wavelength regions for our analysis
is too small to allow a useful statistical study of differences in NAL
properties between subclasses.

\begin{figure}
\centerline{
\includegraphics[height=4in]{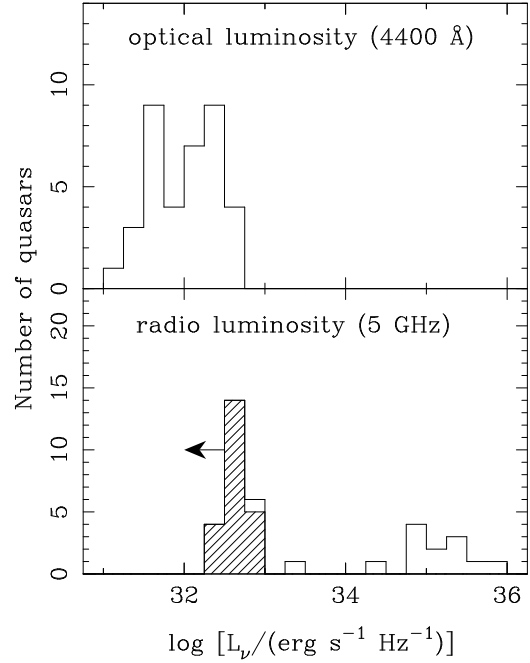}
}
\caption{Distributions of optical (4400~\AA) and radio (5~GHz)
  luminosities of the 37 quasars in our sample. The shaded bins in the
  lower panel correspond to upper limits to the radio flux, as listed
  in Table~\ref{tab:quasars}. Radio-loud objects cluster at 5GHz
  luminosities between $10^{35}$ and $10^{36}$
  ergs~s$^{-1}$~Hz$^{-1}$.\label{fig:lumdist}}
\end{figure}

The properties of the quasars in our sample are summarized in
Table~\ref{tab:quasars} and Figure~\ref{fig:lumdist}. Columns (1) and
(2) of Table~\ref{tab:quasars} give the quasar name and emission
redshift, and columns (3) and (4) the $V$ and $R$ magnitudes. The
optical flux density at 4400~\AA\ in the rest frame is given in column
(5). Column (6) lists the observed radio flux at the frequency of
column (7), if a radio source was detected within 10$^{\prime\prime}$
of the optical source (e.g., Kellermann 1989).  The derived radio flux
at 5~GHz is listed in column (8). Column (9) gives the radio-loudness
parameter, $\rl$, based on which the quasars are labeled as L
(radio-loud) or Q (radio-quiet) in column (10).  The distributions of
optical (at 4400~\AA) and radio (at 5~GHz) luminosities are plotted in
Figure~\ref{fig:lumdist}.

\section{Sample of NAL Systems\label{sec:sample}}

To construct our sample of NAL systems for statistical analysis, we
first examine the 37 quasar spectra, and mark all absorption features
that are detected at a confidence level greater than 5$\sigma$, i.e.,
$[1-R_{c,obs}]/\sigma(R_{c,obs}) \geq 5$ ($R_{c,obs}$ is the observed
residual intensity at the center of a line in the normalized spectrum
and $\sigma(R_{c,obs})$ is its uncertainty).  Next, we identify
\ion{N}{5}, \ion{C}{4}, and \ion{Si}{4} doublets in the following
regions around the corresponding emission line:

\begin{description}

\item [\it \ion{N}{5} absorption doublets:] from $-5,000~\kms$ to
0~\kms\ around the \ion{N}{5} emission line in the spectra that
include this line.\footnote{The velocity offset is defined as negative
for NALs that are blueshifted from the quasar.}  If the spectrum
extends redward of the \ion{N}{5} emission line, we also search for
absorption doublets up to $+10,000~\kms$ to the red of the line. The
velocity range is relatively narrow for this transition because of
contamination from the \lya\ forest, which is rather serious at the
redshift of our target quasars.

\item[\it \ion{C}{4} absorption doublets:] from $-70,000~\kms$ to
0~\kms\ around the \ion{C}{4} emission line in the spectra that
include this line.  If the spectrum extends redward of the \ion{C}{4}
emission line, we also search for absorption doublets up to
$+10,000~\kms$ to the red of the line.

\item[\it \ion{Si}{4} absorption doublets:] from $-40,000~\kms$ to
0~\kms\ around the \ion{Si}{4} emission line in the spectra that
include this line.  If the spectrum extends redward of the \ion{Si}{4}
emission line, we also search for absorption doublets up to
$+10,000~\kms$ to the red of the line.

\end{description}

\noindent
In Table~\ref{tab:quasars}, we list the red limit of the velocity
window that we searched for each quasar.

Absorption troughs that are separated by non-absorbed regions are
considered to be separate lines, even if they are very close to each
other.  The equivalent width is measured for each separate line by
integrating across the absorption profile.  In the case of doublets,
we represent the equivalent width by that of the stronger (blueward)
member.  In total, 261 \ion{C}{4}, 13 \ion{N}{5}, and 92 \ion{Si}{4}
doublets are identified in this manner. To facilitate detailed studies
of the systems, we also searched for single metal lines at the same
redshifts as the doublet lines even if these were in the \lya\
forest. For the statistical analysis, we construct a complete sample
that contains only doublet lines whose stronger members would be
detected even in our most noisy quasar spectrum. We convert the
absorption-feature detection criterion given above,
$[1-R_{c,obs}]/\sigma(R_{c,obs}) \geq 5$, to an equivalent width limit
using equation (1) of Misawa et al. (2003) and we select absorption
features above a certain rest-frame equivalent width. The rest-frame
equivalent width limits are set by the spectrum of the quasar
Q1330$+$0108 as follows: $W_{min}\;$(\ion{C}{4})$\;=0.056$~\AA\ at
$\lambda \approx 5800$~\AA, $W_{min}\;$(\ion{N}{5})$\;=0.038$~\AA\ at
$\lambda \approx 5700$~\AA, and
$W_{min}\;$(\ion{Si}{4})$\;=0.054$~\AA\ at $\lambda \approx
5800$~\AA. These limits apply to the stronger member of each doublet.
The ``homogeneous'' NAL sample defined by these limits contains 138
\ion{C}{4}, 12 \ion{N}{5}, and 56 \ion{Si}{4} doublets.

In order to evaluate the column densities ($N$ in \cmm) and Doppler
parameters ($b$ in \kms) of NALs, we need to deblend the absorption
profiles into narrower components. Thus, we used the software package
{\sc minfit} (Churchill \& Vogt 2001) to fit absorption lines with
Voigt profiles. In the fitting process, we consider the coverage
fraction (defined and discussed in detail in \S\ref{sec:parcov},
below) as a free parameter, as well as the redshift, column density,
and Doppler parameter, and we convolve the model with the instrumental
profile before comparing with the data. Kinematic components are
dropped if a model with fewer components provides an equally
acceptable fit to the data. With this procedure, the NALs of the
homogeneous sample are deblended into 483 \ion{C}{4}, 41 \ion{N}{5},
and 182 \ion{Si}{4} components. The observed profiles of \ion{C}{4},
\ion{N}{5}, and \ion{Si}{4} NALs, best-fitting models, and 1$\sigma$
error spectra, are shown on a velocity scale relative to the flux
weighted line center in the figures of the Appendix.  Almost all the
NALs are deblended into multiple components.

\begin{figure}
\centerline{
\includegraphics{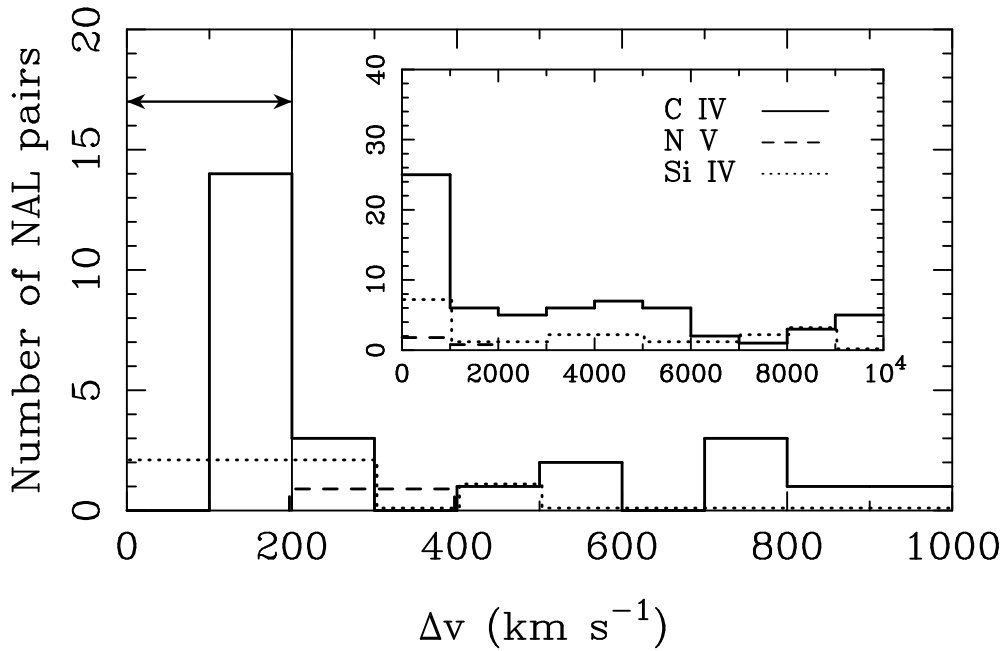}
}
\caption{Two point correlation functions for \ion{C}{4} (solid line),
  \ion{N}{5} (dashed line), and \ion{Si}{4} (dotted line) NALs in
  velocity offset space.  Strong clustering is seen at $\Delta v$
  $\leq$200~\kms\ (indicated by arrows in the main panel), especially
  for \ion{C}{4} NALs. Although we count only the number of pairs
  which are just next to each other, we get a similar result if we
  count all combinations.\label{fig:clustering}}
\end{figure}

To refine our sample for statistical analysis, we combine NALs that
lie within 200~\kms\ of each other into a single system (a so-called
``Poisson system''). This method is based on the assumption that
clustered lines are not physically independent (e.g., Sargent,
Boksenberg, \& Steidel 1988). We chose this clustering velocity after
constructing the distribution of velocity separations between
\ion{C}{4} NALs, which we show in Figure~\ref{fig:clustering}. We note
that our adopted clustering velocity differs from the value of
1,000~\kms\ adopted in previous papers (e.g., Sargent et al. 1988;
Steidel. 1990; Misawa et al. 2002).  We obtain 150 Poisson systems in
this manner, of which 124, 12, and 50 systems contain \ion{C}{4},
\ion{N}{5}, and \ion{Si}{4} NALs, respectively. All NALs are also
classified as AALs and non-AALs. We found 18 associated Poisson
systems, of which 9, 12, and 4 systems contain \ion{C}{4}, \ion{N}{5},
and \ion{Si}{4} NALs, respectively.

Hereafter, we will use the terms ``(Poisson) system'' for a group of
NALs within 200~\kms\ each other, ``line'' or ``doublet'' for an
individual NAL (that may lie within a Poisson system), and
``component'' for a narrow kinematic component described by a single
Voigt profile (as deblended by {\sc minfit})\footnote{As we noted in
\S\ref{sec:sample}, a ``line'' is separated from its neighbors by
non-absorbed regions.}. In Table~\ref{tab:census} we list the total
numbers of systems, lines, and components that we have found for each
of the three strong absorption lines of interest, namely \ion{C}{4},
\ion{Si}{4}, and \ion{N}{5}.

We compute the flux-weighted center of a line (or a Poisson system) as
the first moment of the line profile, i.e.,
\begin{equation}
  \lambda_{obs} = \langle\lambda\rangle =
  \frac{\sum\lambda_i\left[1-R(\lambda_i)\right]}
       {\sum\left[1-R(\lambda_i)\right]}\; ,
  \label{eqn:lambdaobs}
\end{equation}
and the line dispersion, $s$, via
\begin{equation}
  s_{\lambda}=\sqrt{\langle\lambda^{2}\rangle-\langle\lambda\rangle^{2}}\; ,
\end{equation}
where $\langle\lambda^{2}\rangle$ is the second moment of the line
profile, given by
\begin{equation}
  \langle\lambda^{2}\rangle=
  \frac{\sum\lambda_i^{2}\left[1-R(\lambda_i)\right]}
       {\sum\left[1-R(\lambda_i)\right]}\; ,
\end{equation}
The above sums are evaluated over the entire width (full width at zero
depth) of a line.  The quantity $R(\lambda_i)$ is the intensity at
pixel $i$ in the normalized spectrum and $\lambda_i$ is the wavelength
of that pixel.  Using the above quantities, we obtain the
flux-weighted line width from
\begin{equation}
  \sigma(v) = \sqrt{2}\;s_v 
  = \sqrt{2}\left(\frac{s_{\lambda}}{\langle\lambda\rangle}\right)\; c
  \hspace{5mm} {\rm (km}\ {\rm s}^{-1}{\rm )},
  \label{eqn:bpar}
\end{equation}
Using the flux-weighted wavelength of a line we compute its
flux-weighted redshift, and then its velocity offset by the
relativistic Doppler formula, i.e.,
\begin{equation}
  \beta\equiv \frac{\voff}{c}=
  -\frac{(1+z_{em})^2-(1+z_{abs})^2}{(1+z_{em})^2+(1+z_{abs})^2}\;,
  \label{eqn:deltav}
\end{equation}
where $z_{em}$ and $z_{abs}$ are the emission redshift of the quasar
and the absorption redshift of the line, respectively.  The
wavelengths, redshifts, velocity offsets, and flux-weighted widths of
detected lines, computed as described above, are given in the
Appendix, where we also include NALs that do not meet the rest-frame
equivalent width criterion for our homogeneous sample (i.e., they have
$W_{rest} < W_{min}$).

\subsection{Partial Coverage Analysis\label{sec:parcov}}

We identified intrinsic NAL candidates by looking for absorption
troughs that were diluted by unocculted light from the background
source. The optical depth ratio of doublet lines, such as \ion{C}{4},
\ion{Si}{4}, or, \ion{N}{5}, sometimes deviates from the value of 2:1
expected from atomic physics. This discrepancy can be explained if the
absorber covers the background flux source only partially, and the
unabsorbed flux changes the relative depth of the lines (e.g., Wampler
et al. 1995; Barlow \& Sargent 1997; Hamann et al. 1997a; G99).
However, there are other possible explanations, such as local emission
by the absorbers (Wampler, Chugai, \& Petitjean 1995) and scattering
of background photons into our line of sight (G99).  We assume the
simplest model, in which the absorber has a constant optical depth
across the projected area (i.e., a homogeneous model). Then, the
observed intensity as a function of velocity from the line center in
the {\it normalized} spectrum is given by
\begin{equation}
  R(v)=\left[1-C_f(v)+C_f(v)\;e^{-\tau(v)}\right]\; ,
  \label{eqn:normint}
\end{equation}
where $C_f(v)$ is the fraction of background light occulted by the
absorber (hereafter, the ``coverage fraction''), and $\tau(v)$ is the
optical depth at velocity $v$. If a NAL system consists of many
kinematic components (the most common case), these are combined by
multiplying the individual Voigt profiles after making the appropriate
adjustments for partial coverage, following
equation~(\ref{eqn:normint}). Thus we describe the final {\it
normalized} residual intensity by the product
\begin{equation}
  R(v)=\prod_k
  \left[1-C_{f,k}(v)+C_{f,k}(v)\; e^{-\tau_k(v)}\right],
  \label{eqn:product}
\end{equation}
where the index $k$ labels different kinematic components. This
assumption is quite safe if the components do not overlap
significantly in velocity, and is still a good approximation if one
component dominates at each wavelength.

If we measure the optical depth ratio of doublet lines with oscillator
strength values of 2:1 (e.g., \ion{C}{4}, \ion{N}{5}, and \ion{Si}{4})
by fitting their profiles, we can evaluate the coverage fraction as a
function of velocity across the line profile as
\begin{equation}
  C_f(v)=\frac{\left[1-R_r(v)\right]^{2}}{1+R_b(v)-2R_r(v)},
  \label{eqn:cf}
\end{equation}
where $R_{r}$ and $R_{b}$ are the continuum normalized intensities of
the weaker (redder) and stronger (bluer) members of the doublet (see,
for example, Barlow \& Sargent 1997).  G99 refined this technique by
considering two background sources: the continuum source and the broad
emission line region (hereafter BELR).  In this composite picture the
total coverage fraction can be expressed as a weighted average of the
coverage fractions of the two regions, namely,
\begin{equation}
  C_{f}(v)=\frac{C_{c}(v)+w(v)\; C_{BELR}(v)}{1+w(v)},
  \label{eqn:wcf}
\end{equation}
where $C_{c}(v)$ and $C_{BELR}(v)$ are the coverage fractions of the
continuum source and the broad emission line region, and $w(v)$ is the
ratio of the broad line flux to the continuum flux at a given pixel in
the absorption-line profile.

It is quite possible for the absorber to have different optical depths
along different paths (i.e., an inhomogeneous model; deKool, Korista,
\& Arav 2002). Nonetheless, by exploring this inhomogeneous model,
Sabra \& Hamann (2005) found that the average optical depths derived
from homogeneous and inhomogeneous models are consistent with each
other within a factor of $\leq$1.5, unless a small fraction of the
projected area of the absorber has a very large optical
depth. Moreover, in most cases, one can find acceptable fits to the
absorption profile using a simple combination of just a few Voigt
components. Therefore, we adopt the homogeneous optical depth model,
as described by equation~(\ref{eqn:cf}), with the understanding that
$C_f$ describes the fraction of {\it all} background photons that pass
through the absorber. We consider the composite picture described by
equation~(\ref{eqn:wcf}), only if it critically affects our
conclusions.

We use two methods to evaluate \cf; (i) the {\it pixel-by-pixel}
method (e.g., G99), in which we apply equation~(\ref{eqn:cf}) to each
pixel in the normalized spectrum, and (ii) the {\it fitting method}
(e.g., Ganguly et al. 2003) in which we fit the absorption profiles
using {\sc minfit}, treating $C_{f}$ as well as $\log N$, $b$, and
$z_{abs}$ as a free parameter. In the former case we obtain a value of
\cf\ for each pixel in the line profile, while in the latter case we
obtain a value of \cf\ for each kinematic (Voigt) component.  Our
derived \cf\ values are subject to a number of additional
uncertainties resulting from line blending, uncertainties in the
continuum fit, the convolution of the spectrum with the line spread
function (LSF) of the spectrograph (G99; only applies to the
pixel-by-pixel method), and Poisson errors in regions with weak lines
(only applies to the fitting method). The methodology for assessing
these uncertainties was developed in M05 with the help of extensive
simulations. We do not use spectral regions that are strongly affected
by these uncertainties, and we consider these sources of error when
evaluating whether \cf\ deviates from unity, as described below. The
\cf\ values evaluated by both of the above methods are overplotted on
the observed spectrum of each doublet shown in
Figure~\ref{fig:pcovplots} of the Appendix.

The {\sc minfit} routine sometimes gives unphysical coverage fractions
such as $\cf<0$ or $\cf>1$. Tests by M05 showed that coverage
fractions produced by {\sc minfit} are very sensitive to continuum
level errors, especially for very weak components whose \cf\ values
are close to 1, which suggests a cause for these unphysical results.
In such cases, other fit parameters, such as the column density and
Doppler parameter, have no physical meaning. Therefore we re-evaluate
them assuming $\cf=1$ (with no error bar) for those components, and by
refitting the rest of the line.

Based on the results of the coverage fraction analysis, we separate
all NALs into three classes based on the reliability of the conclusion
that they are intrinsic: classes A and B respectively contain
``reliable'' and ``possible'' partially covered NALs (i.e., intrinsic
NAL candidates), and class C consists of NALs with no evidence for
partial coverage (i.e., intervening or unclassified NALs).  We may
regard classes A--C as including progressively smaller fractions of
intrinsic NALs and we will treat them as such in our subsequent
statistical analysis.

\begin{figure}
\centerline{
\includegraphics[height=8in]{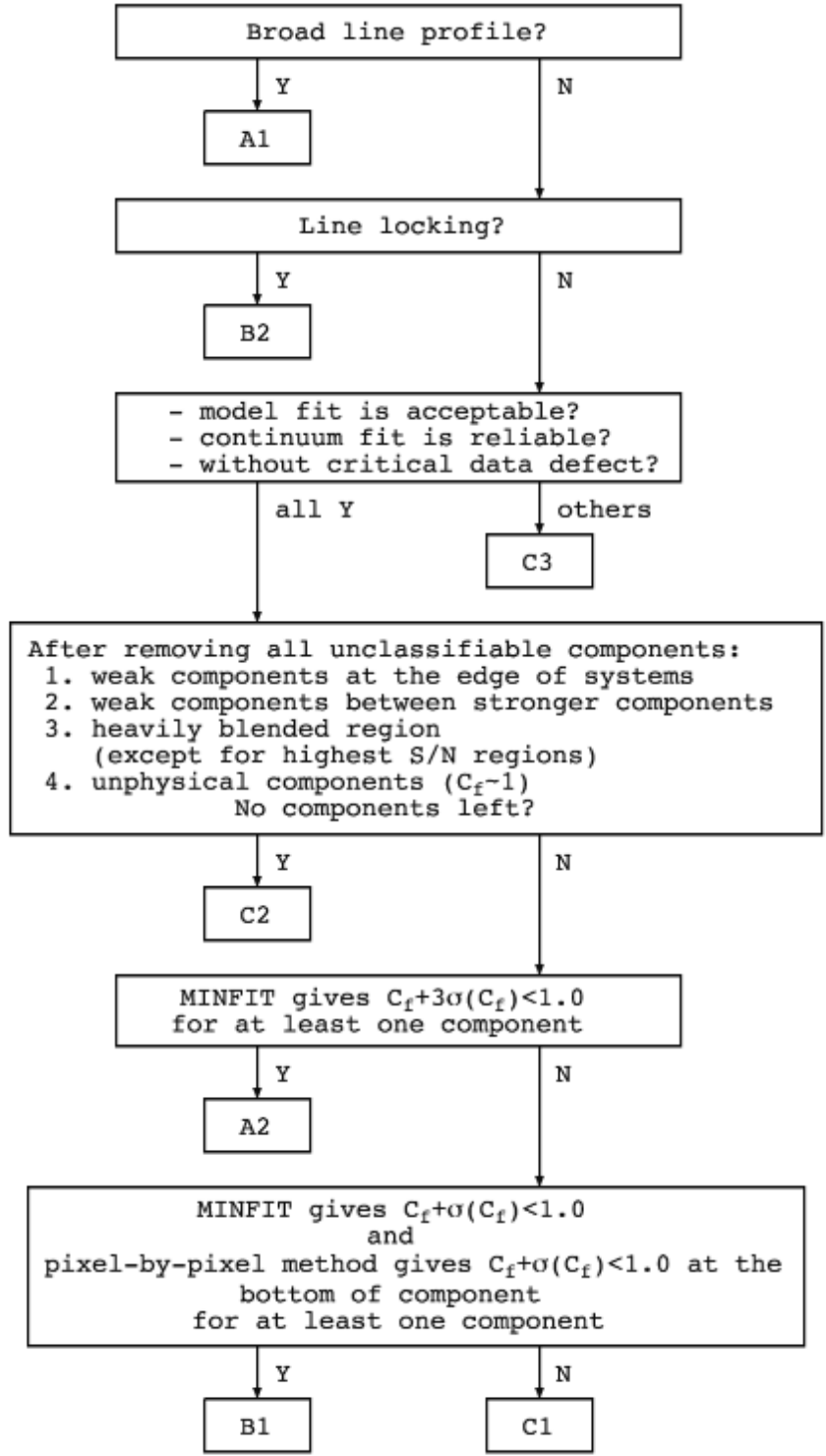}
}
\caption{Flow chart illustrating our scheme of NAL classification. The
  classes indicate how reliably we can determine whether the NALs are
  intrinsic or not, based on partial coverage analysis. A more
  detailed discussion of this scheme is given in \S\ref{sec:parcov} of
  the text.\label{fig:flow}}
\end{figure}

We classified four mini-BALs as class A systems. We also classified
three NALs into class B, although they did not exhibit partial
coverage, because they show line-locking, which is often seen in
intrinsic NALs at $\zabs \sim\zem$. In the list below and in the flow
chart of Figure~\ref{fig:flow}, we give the criteria and method that
we used to classify NALs into reliability classes.

\begin{description}
 \item[Class A:]Reliable intrinsic NAL candidates:
  \begin{description}
      \item[A1:]smooth and broad (self-blended) line profile (i.e.,
                 mini-BAL).
      \item[A2:]{\sc minfit} gives \cf $+$ 3$\sigma$(\cf) $<$ 1 for at
                 least one component.
  \end{description}
 \item[Class B:]Possible intrinsic NAL candidates:
  \begin{description}
      \item[B1:]Both {\sc minfit} and the pixel-by-pixel method give
                 \cf $+$ $\sigma$(\cf) $<$ 1 at the center of a
                 component for at least one kinematic component.
      \item[B2:]line-locked.
  \end{description}
 \item[Class C:]Lines without partial coverage and unclassifiable
                 lines:
  \begin{description}
      \item[C1:]{\sc minfit} gives $\cf+\sigma(\cf)\geq 1$ for all
                 components.
      \item[C2:]No components can be used for classification because
                 they suffer from systematic errors as described in
                 the next paragraph.
      \item[C3:]The system is not acceptable for classification
                 because of problems due to model fit, continuum fit,
                 or critical data defect.
  \end{description}
\end{description}

During line classification, we ignore components that satisfy the
following criteria. These were established by following the same
procedure to fit strong \ion{Mg}{2} doublets identified in the same
spectra.  For the strong \ion{Mg}{2} lines we would expect full
coverage for all components, but we find \cf\ to be inconsistent with
unity if these criteria apply:

\begin{enumerate}

\item 
Weak components at the edge of a system.

\item 
Weak components between much stronger components.

\item 
Components in a heavily blended region (except for regions with
extremely high S/N).

\item 
Components for which we obtained unphysical values of \cf\ in the
first fitting trial. As we noted above, we set $\cf=1$ for such
components and repeat the fit for the other components in the system.

\end{enumerate}

If at least one component of a NAL satisfies one of the criteria for
class A or B, we would classify the NAL as an intrinsic NAL candidate,
even if the other components are consistent with full coverage. This
applies even for a Poisson system for which only one component in one
of the separate lines satisfies the class A or B criteria. Also, if a
Poisson system has more than one doublet transition detected and one
or more of the transitions (\ion{C}{4}, \ion{N}{5}, or \ion{Si}{4})
belongs in class A, it is considered a class A intrinsic system.
Similarly, such a system will be considered a class B system if one or
more transitions fall in class B (and none is class A).

Placing a Poisson system in class A or B if only one component
satisfies the classification criteria involves the following caveat:
the class A or B NAL may be superposed on an intervening system by
chance. We have thus evaluated the probability of this chance
superposition, using the information in Table~\ref{tab:stats}. For
example, we have searched for {\it associated} \ion{C}{4} NALs in a
velocity range $\delta\beta=0.17$ and found 3 class A+B and 6 class C
NALs. Since each NAL typically spans a velocity window of 200~\kms\ or
less, the class C NALs cover 2.4\% of this velocity window. Therefore,
the probability that any one of the 3 class A+B NALs would be blended
with a class C NAL (i.e., the expected number of blends) is 0.07.
Similarly, we estimate the number of blends to be 0.25 in the {\it
non-associated} \ion{C}{4} velocity range, 0.03 in the \ion{N}{5}
velocity range (associated only), and 0.07 in the \ion{Si}{4} velocity
range (non-associated only). Thus, we conclude that the danger of a
blend is negligible.

In our final sample of 150 Poisson NAL systems, 28 fall into class A,
11 into class B, and 111 into class C. The number of systems, lines,
and components are summarized in Table~\ref{tab:census}.  In the
Appendix, we list the detailed classification results, along with the
subcategory numbers that give the basis for the classification (e.g.,
A2, C3, ...). In the Appendix, we also give detailed comments on
individual NAL systems. In Table~\ref{tab:census}, we give a census of
the resulting Poisson systems, NALs and kinematic components.

\section{Results\label{sec:results}}

Using the complete NAL sample constructed in the previous section, we
first investigate the density of NALs per unit redshift interval and
per unit velocity interval (i.e., the velocity offset distribution,
$dN/dz$ and $dN/d\beta=c\; dN/dv$, respectively). Next, we examine the
relative numbers of {\it intrinsic} NALs in different
transitions. This is of interest because these relative numbers could
be indicative of the ionization structure of the absorbers and also of
their locations relative to the continuum source.  Our analysis
provides only lower limits on the intrinsic NAL fractions, because
intrinsic absorber need not always to show partial coverage, and
because some of our sample quasar spectra cover only very small
redshift windows for NAL detection. We also examine the distributions
of coverage fractions and line profile widths, and consider the
ionization conditions of the intrinsic NALs in our sample.  Finally,
we briefly compare the properties of quasars and the properties of
intrinsic NALs.

\subsection{Velocity Offset and Equivalent Width Distribution\label{sec:velew}}

\begin{figure}
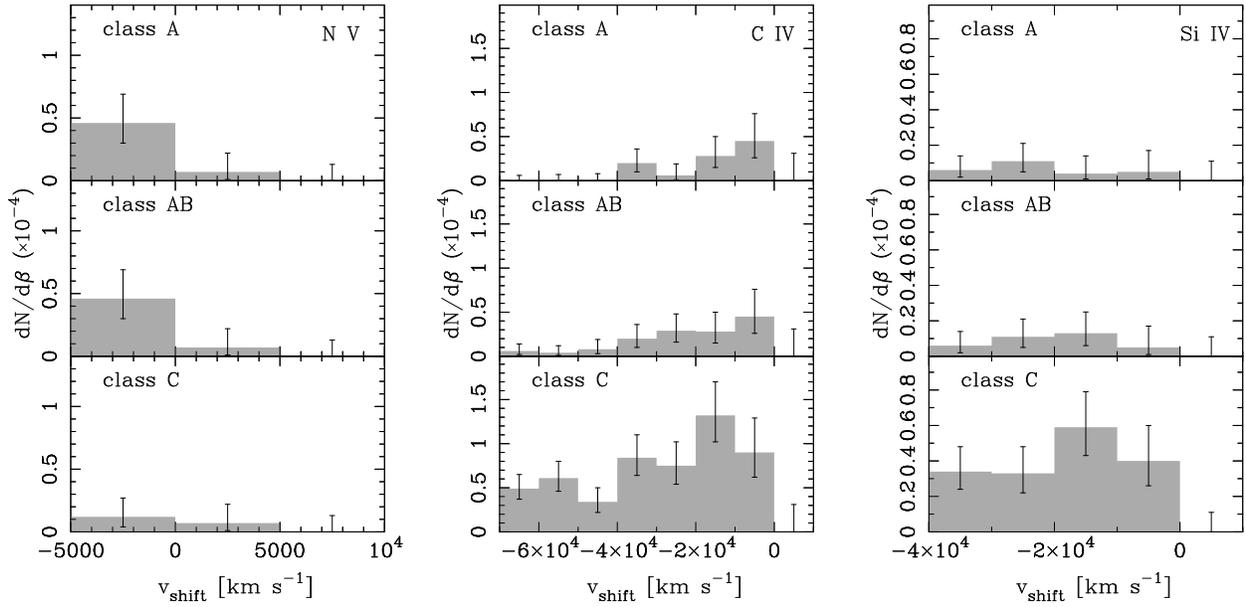

\centerline{
\includegraphics[width=2.2in]{f4a.eps}
\includegraphics[width=2.2in]{f4b.eps}
\includegraphics[width=2.2in]{f4c.eps}
}
\caption{Velocity offset distribution of \ion{C}{4}, \ion{N}{5}, and
  \ion{Si}{4} systems. For each transition we plot separately the
  distribution of class A systems (reliable), class A+B systems
  (reliable+possible), and class C systems
  (intervening/unclassified).\label{fig:voffdist}}
\end{figure}

In evaluating $dN/dz$, we exclude segments of spectra in echelle order
gaps, which affect about 8.8\%, 16.0\%, and 23.0\%\ of the spectral
regions at $\sim$5,500~\AA, $\sim$6,000~\AA, and $\sim$6,500~\AA,
respectively.  We implicitly assume that all NALs arise in outflows
from the quasars and we use the velocity offset computed by
equation~(\ref{eqn:deltav}). In Table~\ref{tab:stats}, we summarize
our derived values of $dN/dz$ and $dN/d\beta$ for different
transitions and for different velocity offsets relative to the quasar,
and we also break down the results by reliability class.  The velocity
offset distributions of \ion{C}{4}, \ion{N}{5}, and \ion{Si}{4} NALs
are presented in Figure~\ref{fig:voffdist}, broken down according to
reliability class.  It is noteworthy that intrinsic NAL candidates
(i.e., class A and B) are found not only among AALs but also among
non-AAL systems.

In previous studies using low resolution spectra, the velocity offset
distributions of \ion{C}{4} NALs are found to be almost uniform up to
$\voff\sim -70,000~\kms$ (the maximum velocity at which the \ion{C}{4}
doublet can be detected without blending with the \lya\ forest).
However, a significant excess of AALs was found within 5,000~\kms\ of
radio-loud and steep-spectrum quasars (Young, Sargent, \& Boksenberg
1982; Richards et al. 1999, 2001; F86).  Weymann et al. (1979)
presented evidence that the distribution of intrinsic \ion{C}{4} NALs
extends up to $\voff\sim -18,000$~\kms. Richards et al. (1999) found
an excess of \ion{C}{4} NALs with $\voff\gsim -65,000~\kms$ in
optically luminous quasars, in radio-quiet over radio-loud quasars,
and in flat-radio spectrum quasars over steep-radio spectrum
quasars. Our results show no remarkable excess of NALs in radio-loud
quasars, nor any strong excess of NALs near the quasars in either the
radio-loud or quiet subsamples. This is not surprising, however, since
our subsamples of radio-loud and radio-quiet quasars are relatively
small and heterogeneous.

\begin{figure}
\centerline{
\includegraphics[height=4in]{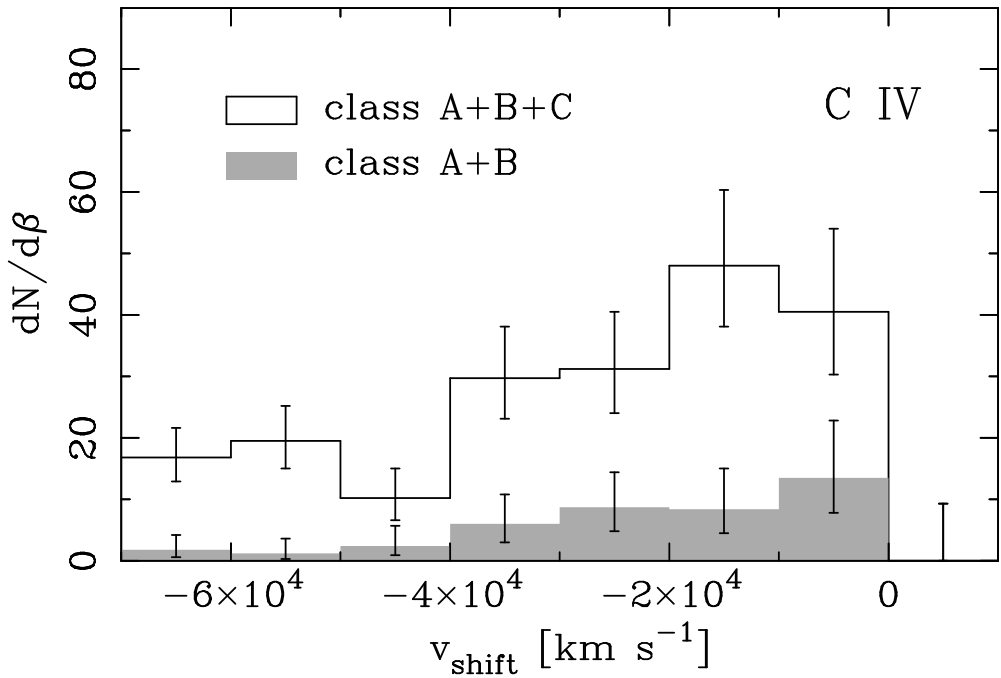}
}
\caption{Velocity offset distribution of all \ion{C}{4} systems
  compared to the distribution of class A+B (reliably and possibly
  intrinsic) \ion{C}{4} systems. 
\label{fig:weym}}
\end{figure}

Considering all \ion{C}{4} NALs (from classes A, B, and C) together,
we do find (see Table~\ref{tab:stats}) that, for \ion{C}{4} NALs the
values of $dN/dz$ and $dN/d\beta$ in associated regions are about
twice as large as those in non-associated regions, in general
agreement with the results of Weymann et al (1979). In
Figure~\ref{fig:weym}, we show the velocity offset distribution of all
\ion{C}{4} NALs from our survey compared to that of class A$+$B NALs.

\begin{figure}
\centerline{ \includegraphics[height=4in]{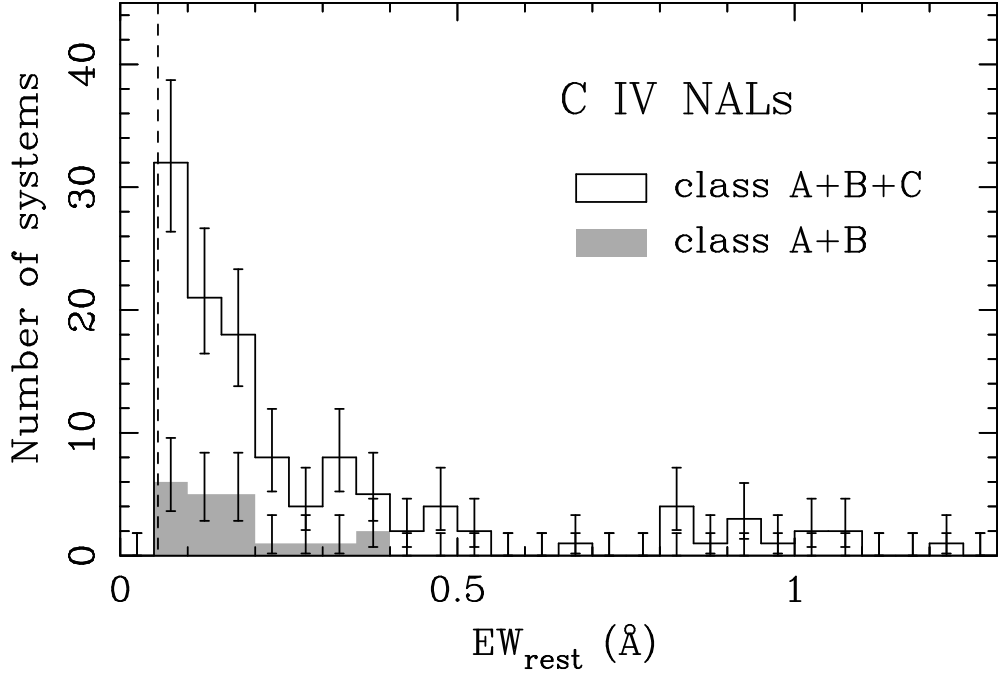} }
\caption{Distribution of rest-frame equivalent width of all \ion{C}{4}
  systems, compared to the distribution of class A+B (reliably and
  possibly intrinsic) \ion{C}{4} NALs. The vertical dashed line marks
  our detection limit of 0.056~\AA. \label{fig:EWdist}}
\end{figure}

The distribution of rest-frame equivalent widths of the \ion{C}{4}
Poisson systems detected in our survey is shown in
Figure~\ref{fig:EWdist}.  This distribution rises sharply towards our
detection limit of 0.056~\AA.  We note that the vast majority of our
NALs are considerably weaker than those included in previous surveys
at low spectral resolution. For example, the survey of Weymann et
al. (1979) has a rest frame equivalent width limit of about 0.6~\AA\
(higher than 87\% of the NALs in our \ion{C}{4} sample) while the
surveys of Young et al. (1982) and F86 have a detection limit of
0.3~\AA\ (higher than 70\% of the NALs in our sample).  The more
recent Vestergaard (2003) study was also conducted using low spectral
resolution, so it focused on \ion{C}{4} absorption lines stronger than
$\sim$0.3~\AA .

A comparison of the equivalent width distributions between our study
and these previous low resolution studies is a useful indication of
possible biases due to quasar properties.  However, such a comparison
must focus only on our stronger systems.  Also, we must group together
different lines that fall within 200~\kms\ of each other in order to
compare with systems found at low resolution (as we did in
Figure~\ref{fig:EWdist}).  First we focus on systems with
$W_r$(\ion{C}{4}) $>$ 1.5~\AA\ since we are concerned about possible
biases due to rejecting BAL quasars from our sample, and since
Vestergaard (2003) targeted the strongest systems as likely intrinsic
candidates.  If we take into account Vestergaard's different method of
measuring the equivalent width of \ion{C}{4}, we find that her sample
includes 6/114 quasars that have systems with $W_r$(\ion{C}{4}) $>$
1.5~\AA .  However, the Vestergaard (2003) sample has approximately
equal numbers of radio-loud and radio-quiet quasars, unlike ours, in
which only 1/3 of the quasars are radio-loud.  If we consider only her
radio-quiet sample, only 1/48 quasars has a \ion{C}{4} NAL with
$W_r$(\ion{C}{4}) $>$ 1.5~\AA .  In her radio-loud sample, 5/66
quasars have very strong \ion{C}{4} NALs.  Thus we would predict that
in our sample of $37$ quasars (with only $12$ radio-loud) we would
find $1.4$ very strong \ion{C}{4} NALs.  Though we did not find any,
our number is certainly consistent with this estimate, and we do find
a system with $W_r$(\ion{C}{4}) $\sim$1.0~\AA, just below that limit.
We note that due to the nature of Vestergaard's study, her sample has
a larger than representative fraction of radio-loud quasars.  Our
sample also has a somewhat larger fraction of radio-loud quasars
(37\%) compared to the general population ($\sim$10\%), which would
lead to a bias towards detecting very strong NALs compared to the
general population.

We next consider somewhat weaker NALs, down to $W_r$(\ion{C}{4}) =
0.3~\AA, to facilitate another comparison with the work of Vestergaard
(2003).  She found that 27\% of the quasars in her sample had
\ion{C}{4} absorption for which the sum of the equivalent widths of
the two members of the doublet were $>0.5$~\AA .  In our sample, we
only have 13 quasars for which the most of the \ion{C}{4} associated
region is covered.  Of those 13 quasars, 9 have $W_r$(\ion{C}{4}) $>$
0.056~\AA\ NALs, but only 1(8\%) has $W_r$(\ion{C}{4}) $>$ 0.3~\AA .
This is somewhat smaller than we would predict (3.5/13 = 27\%) based
on Vestergaard's results, however we again note the differences in the
radio properties between the samples.  Furthermore, we do not have
full coverage of the associated region in all of the 13 quasars, and
only 3 of the 13 are radio-loud.  It is also worth noting that
typically the Vestergaard (2003) quasars are 1-2 magnitudes less
luminous than the quasars in our sample.

\subsection{Line Width Distribution}

\begin{figure}
\centerline{
\includegraphics[height=4in]{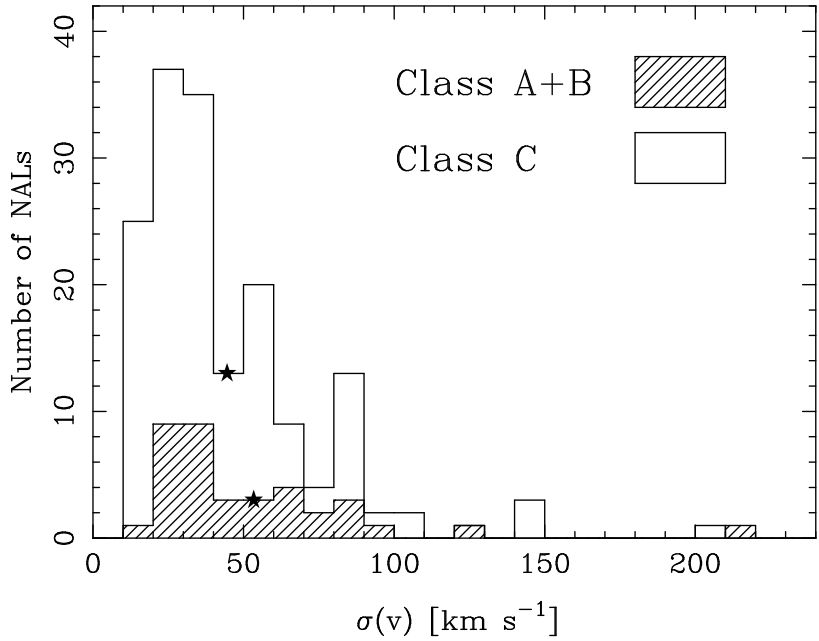}
}
\caption{Distribution of the flux-weighted line width ($\sigma(v)$;
  calculated by eqn. 4) of intrinsic NALs (class A and B) and
  intervening/unclassified NALs (class C). The filled stars indicate
  the average $\sigma(v)$-values for intrinsic NALs (53.4~\kms) and
  intervening NALs (44.5~\kms).\label{fig:bdist}}
\end{figure}

Historically, it has been very difficult to separate intrinsic NALs
from intervening NALs without high resolution spectra because of the
similarities of their line profiles. However, if intrinsic NALs arise
from absorbers similar to those producing BAL and mini-BAL systems
(widths\gsim 2,000~\kms), they could have larger widths compared to
those of intervening NALs. If a tendency for large widths is found
among intrinsic NALs, this may be used as an additional indicator of
their nature.  In Figure~\ref{fig:bdist}, we compare the distribution
of the flux-weighted line width [$\sigma(v)$; see
eqn. (\ref{eqn:bpar})] of intrinsic NALs (classes A and B) to that of
intervening/unclassified NALs (class C). We do not see any substantial
differences between the distributions that would allow us to
distinguish between intervening and intrinsic NALs. The average value
of $\sigma(v)$ for intrinsic NALs (53.4~\kms) is slightly larger than
that for intervening NALs (44.5~\kms). A Kolmogorov-Smirnov test gives
a probability of 12\% that the two distributions have been drawn by
chance from the same parent population.

\subsection{Relative Numbers of Intrinsic NALs in Different Transitions}

The numbers of \ion{C}{4}, \ion{N}{5}, and \ion{Si}{4} NALs in each of
the coverage fraction classes are listed in
Table~\ref{tab:stats}.\footnote{We considered also subsamples of
radio-loud and radio-quiet quasars but found no significant
differences from the total sample.}  We also list in
Table~\ref{tab:stats} the values of $dN/dz$ and $dN/d\beta$ for each
type of system. Based on these values, reliable (class A) intrinsic
NALs make up 11\% of all \ion{C}{4} NALs, 75\% of all \ion{N}{5} NALs,
and 14\% of all \ion{Si}{4} NALs.  These fractions increase to 19\%,
75\%, and 18\%\ after adding possible intrinsic NALs (class
B). Concentrating on different transitions in turn, we note the
following:

\begin{description}

\item[\it \ion{N}{5}. ---] The above fractions of \ion{N}{5} NALs that
are intrinsic refer only to the AAL velocity range ($\voff \leq
-5,000~\kms$) because contamination by the \lya\ forest prevented us
from searching for NALs at larger velocity offsets. Nonetheless, it is
worth noting that 3/4 of these associated \ion{N}{5} NALs are
intrinsic.  The density of intrinsic NALs in the associated region is
thus quite high ($dN/dz=3.4$ and $dN/d\beta=13.7$).

\item[\it \ion{C}{4}. ---] The fraction of intrinsic \ion{C}{4} NALs
in associated regions (33\%) is slightly higher than the fraction we
found in non-associated regions (10\%--17\%). If the fraction of
intrinsic NALs at $\voff>-5,000~\kms$ was actually the same as the
fraction at $\voff<-5,000~\kms$, the probability that 3 or more out of
9 \ion{C}{4} AALs are intrinsic to the quasar (based on the binomial
distribution) is 5.1\% .  If we consider only the radio-quiet
subsample, the probability is even smaller, 1.2\%. In the radio-loud
sample, we found no intrinsic \ion{C}{4} AALs, but this does not mean
that there is an actual deficiency of intrinsic \ion{C}{4} AALs,
because our 12 spectra of radio-loud quasars have only a small
redshift path for detecting \ion{C}{4} AALs ($\delta z=0.64$).

\item[\it \ion{Si}{4}. ---] We found no intrinsic \ion{Si}{4} AALs in
the redshift window of $\delta z=1.26$ that we searched.

\end{description}

\subsection{Fraction of Quasars Showing Intrinsic NALs}

To investigate the geometry of absorbing gas around the quasars we
count the number of intrinsic NALs per quasar, and we evaluate the
fraction of quasars hosting intrinsic NALs. These values constrain the
global covering factor ($\Omega/4\pi$)\footnote{Here it is important
to make the distinction between the terms ``coverage fraction'' and
``global covering factor''.  The former term (defined in
\S\ref{sec:parcov}) refers to the fraction of the projected area of
the background source occulted by an absorbing parcel of gas or,
equivalently, to the fraction of background photons that pass through
the absorber. The latter term refers to the total solid angle
subtended by the ensemble of absorbers around the source.}  of the
continuum source and BELR by the absorber and the distribution of
absorbing gas around the quasar central engine.  At $z < 1$, G01 found
that radio-loud, flat-spectrum quasars with compact radio morphologies
(i.e., with a face-on accretion disk) lack AALs down to a limiting
equivalent width of $W_{rest}=0.34$~\AA.  This suggests that the
presence of intrinsic NALs depends on the inclination of the line of
sight to the symmetry axis of the quasar central engine.

\begin{figure}
\centerline{
\includegraphics[height=4in]{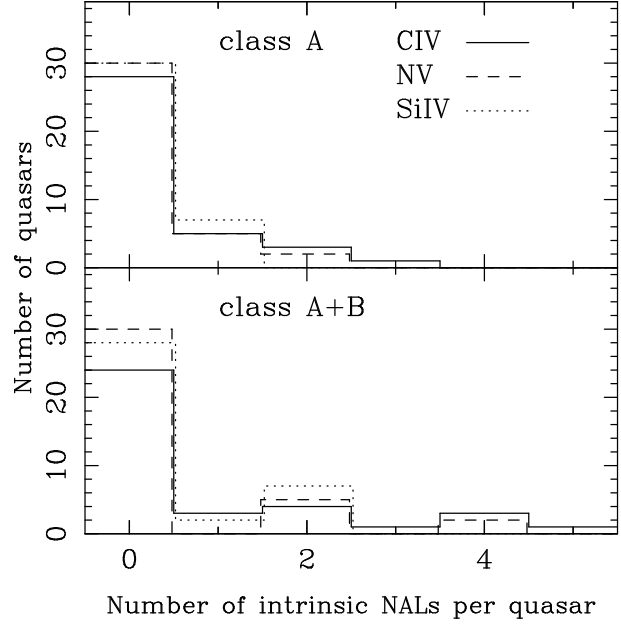}
}
\caption{Distribution of the number of intrinsic systems per
  quasar. The top panel includes only class A (reliable) systems,
  while the lower panel includes class A+B (reliable+possible)
  systems. In each panel we show the distribution of \ion{C}{4}
  systems as a solid line, the distribution of \ion{N}{5} systems as a
  dashed line and the distribution of \ion{Si}{4} systems as a dotted
  line. We note that the number of intrinsic systems per quasars in
  these histogram is strictly a lower limit (see discussion in
  \S\ref{sec:results} of the text).\label{fig:naldist}}
\end{figure}

In Table~\ref{tab:stats} we list the fraction of NALs that are
intrinsic and the fraction of quasars that have intrinsic NALs. We
count only quasars that have velocity windows for NAL detection of
$\delta\beta > 0.01$ ($\delta v > 3,000~\kms$) in the transition of
interest. In Figure~\ref{fig:naldist}, we plot the distribution of the
number of intrinsic NALs per quasar. The upper panel shows the
distribution of only class A NALs, while the lower panel includes both
class A and B NALs.

About 24\% of quasars (9 out of 37) have at least one reliable (class
A) intrinsic \ion{C}{4} NAL. Taking possible (class B) NALs into
consideration, 32\%\ (12 out of 37) of quasars have at least one
intrinsic \ion{C}{4} NAL. The corresponding fractions for \ion{N}{5}
and \ion{Si}{4} (19--24\%) are lower than for \ion{C}{4}. However, we
cannot immediately conclude that the global covering factor of
\ion{C}{4} absorbers is higher than those of \ion{N}{5} and
\ion{Si}{4}, because all of these values are just lower limits since
some intrinsic NALs could have full coverage
\footnote{A parcel of gas in the vicinity of the continuum source
could be large enough to cover the source completely. In such a case
our partial coverage method will yield \cf\ = 1, even though the
absorption lines are really intrinsic.}  and also because we were not
able to search for \ion{N}{5} NALs over a wide range in \voff. If we
consider all three transitions, 43\%\ (16 out of 37) of the quasars
have at least one class A intrinsic NAL (in at least one of the three
transitions) and 54\%\ (20 out of 37) of the quasars have at least one
class A or class B intrinsic NAL.

\subsection{Distribution of Coverage Fractions}

\begin{figure}
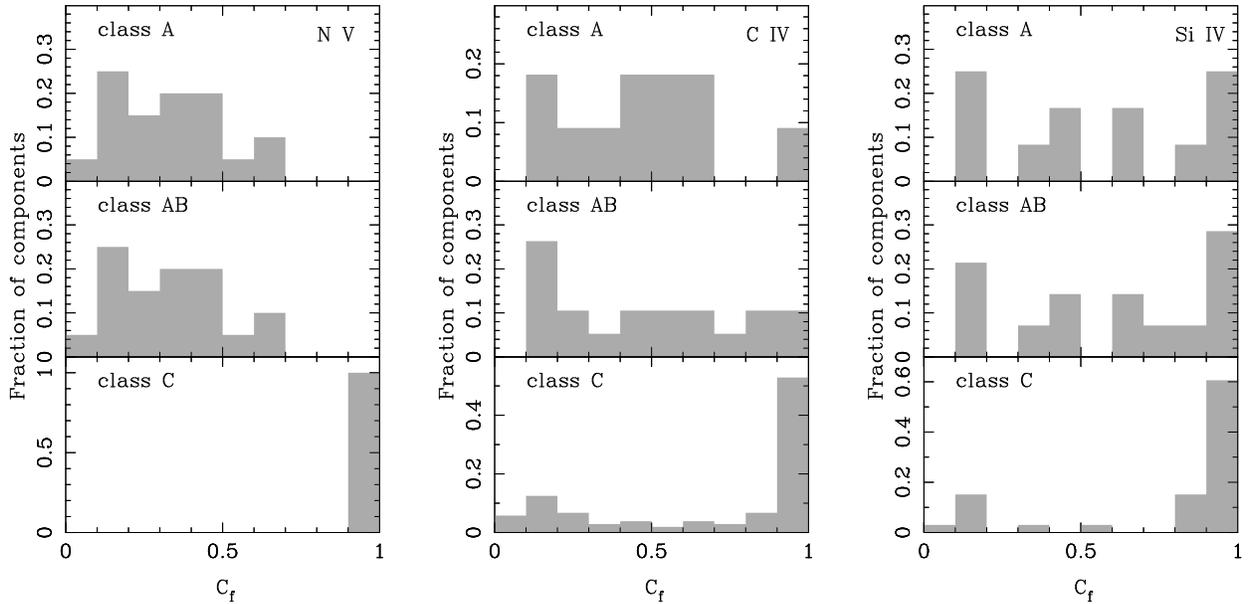

\centerline{
\includegraphics[width=2.2in]{f9a.eps}
\includegraphics[width=2.2in]{f9b.eps}
\includegraphics[width=2.2in]{f9c.eps}
}
\caption{Distribution of coverage fractions of \ion{C}{4}, \ion{N}{5},
  and \ion{Si}{4} NAL components. We plot only components whose \cf\
  values are physical (i.e., $0 < \cf$ and $\cf < 1$) and evaluated
  with high reliability (i.e., $\sigma$(\cf) < 0.1). For each
  transition we plot separately the distribution of class A NALs
  (reliable), class A+B NALs (reliable+possible), and class C NALs
  (intervening/unclassified). Even though some class-C NALs have $\cf
  < 1$. These values are consistent with 1 within their $1\sigma$
  errors.
\label{fig:cfdist}}
\end{figure}

The normalized distribution of the coverage fraction for individual
NAL components is presented in Figure~\ref{fig:cfdist}, where it is
broken down by transition and by reliability class. We plot only
components for which we get physical covering fractions. The mean and
median values of \cf\ in class A$+$B NALs are, respectively, 0.38 and
0.33 for \ion{N}{5}, 0.50 and 0.48 for \ion{C}{4}, and 0.56 and 0.52
for \ion{Si}{4}. Putting together NALs from all three transitions, we
find mean and median values of \cf\ of 0.48 and 0.45,
respectively. Focusing on the \cf\ distribution among \ion{C}{4} NALs
(the largest of all populations shown in Fig.~\ref{fig:cfdist}) we
note that in class A$+$B NALs are distributed uniformly between 0 and
1, while more than half of the class C NALs have coverage fraction
larger than 0.9 (this was one of the criteria for placing a component
in this class).

In addition to the NAL components shown in Figure~\ref{fig:cfdist},
Table~\ref{tab:census} shows that 56\%, 34\%, and 58\%\ of \ion{C}{4},
\ion{N}{5}, and \ion{Si}{4} NAL components in the homogeneous sample
have unphysical measured \cf\ values.  This usually indicates a
doublet ratio slightly greater than two.  If we assume that all
components whose \cf\ values are unphysical or are consistent with
full coverage (i.e., $\cf + \sigma(\cf) \geq 1.0$) really have full
coverage, we find that 73\%\ (355/483), 41\%\ (17/41), and 72\%\
(131/182) of \ion{C}{4}, \ion{N}{5}, and \ion{Si}{4} components cover
the background flux sources.  If we consider only intrinsic systems,
these fractions would be 59\%\ (44/75)), 26\%\ (8/31), and 31\%\
(8/26), respectively.  Generally, we display only components whose
\cf\ values are physical and whose errors are small enough (i.e.,
$\sigma$(\cf) $<$ 0.1) to avoid any uncertainties from \cf\ value
estimation.

For some intrinsic NAL systems, coverage fractions have been
determined for more than one transition (e.g., Ganguly et al. 2003;
Yuan et al. 2002). G99 found that the \ion{C}{4}, \ion{N}{5}, and
\ion{Si}{4} coverage fractions are similar to each other for a few
systems, while Petitjean \& Srianand (1999) and Srianand \& Petitjean
(2000) noted that higher ionization transitions tend to have larger
coverage fraction than lower ionization transitions. This type of
effect could occur if the sizes of the background sources (i.e.,
continuum source, BELR) and/or the absorber are not the same for all
transitions.

\begin{figure}
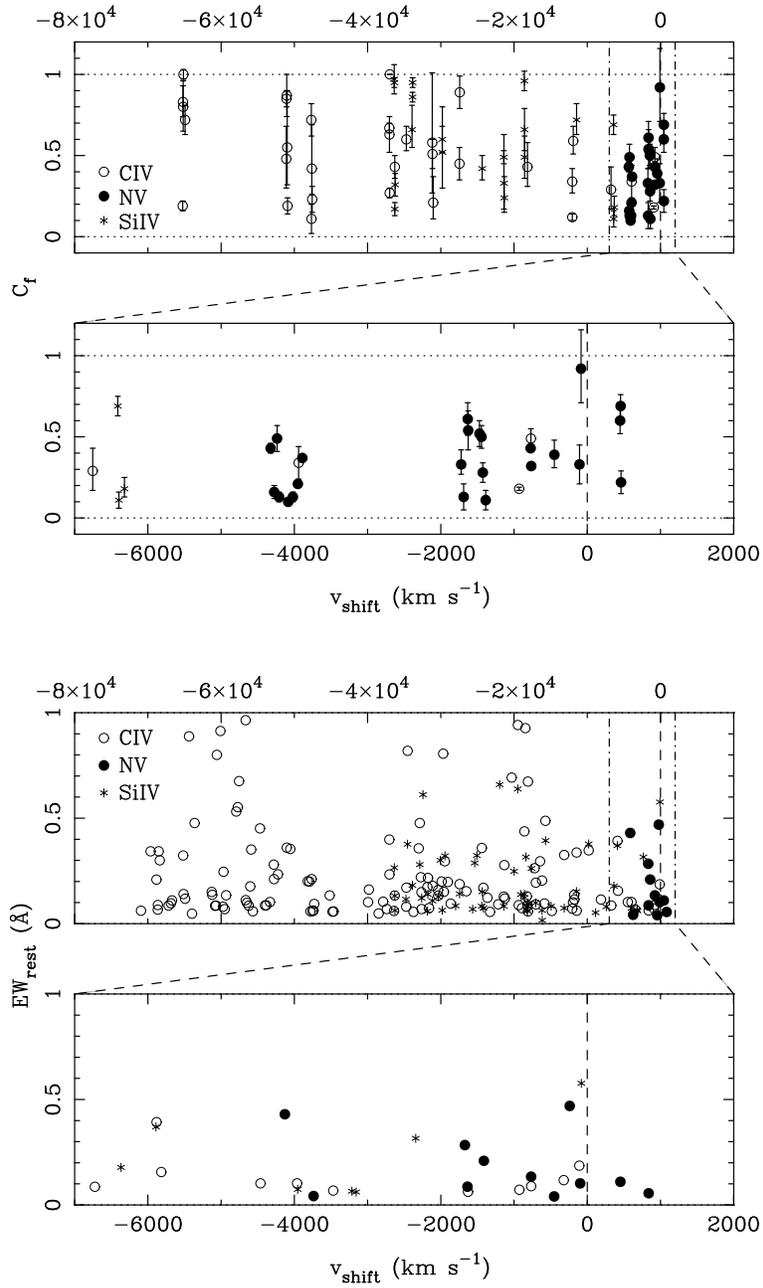

\centerline{\includegraphics[height=3.5in]{f10a.eps}}
\centerline{\includegraphics[height=3.5in]{f10b.eps}}
\caption{Relationship between the velocity offset of intrinsic NALs
  (class A+B) from the quasars and their coverage fraction (top) and
  rest-frame equivalent width (bottom). In the top figure, we plot
  only components whose \cf\ values are physical (i.e., $0 < \cf$ and
  $\cf < 1$) and evaluated with high reliability (i.e., $\sigma$(\cf)
  $<$ 1.0). The upper panel in each case presents a wide velocity region
  (from $-80,000$ to 10,000~\kms). The lower panel focuses on the
  narrower velocity window from $-7,000$ to 2,000~\kms. Open circles
  denote \ion{C}{4} NALs, filled circles denote \ion{N}{5} NALs, and
  asterisks denote \ion{Si}{4} NALs. \label{fig:cfvoff}}
\end{figure}

Unfortunately, our results do not allow a direct comparison between
different transitions within the same system because (a) detecting
multiple transitions from the same system was rather rare and (b)
multiple components in the same intrinsic NAL often have different
\cf\ values, preventing us from assigning a single \cf\ value to a NAL
or a system.  Therefore, we compare the overall distributions of the
physically meaningful (between 0 and 1) \cf\ values of the 34
\ion{C}{4}, 25 \ion{N}{5}, and 20 \ion{Si}{4} components found in 23
\ion{C}{4}, 9 \ion{N}{5}, and 9 \ion{Si}{4} class A and B intrinsic
NALs, in Figure~\ref{fig:cfdist}. An alternative illustration of these
distributions is shown in Figure~\ref{fig:cfvoff} where we plot the
coverage fraction of each NAL component against its offset velocity. A
visual inspection of these two figures suggests that the associated
intrinsic \ion{N}{5} NALs prefer smaller values of \cf\ than the
\ion{C}{4} and \ion{Si}{4} intrinsic NALs.  Nearly all \ion{N}{5}
components show coverage fractions less than 0.5, while the coverage
fractions of \ion{C}{4} and \ion{Si}{4} NALs range from nearly 0 up to
nearly 1. This difference is also reflected in the mean and median
values of \cf\ reported above. However, a Kolmogorov-Smirnov test
yields a probability of 17\% that the \ion{C}{4} and \ion{N}{5} \cf\
distributions were drawn from the same parent population. Similarly,
the probability that the \ion{N}{5} and \ion{Si}{4} \cf\ distributions
were drawn from the same parent population is 7\%. Larger samples are
needed to reach a definitive conclusion.

\subsection{Other Relations between Intrinsic NAL and Quasar Properties}

\begin{figure}
\centerline{
\includegraphics[angle=-90,width=7in]{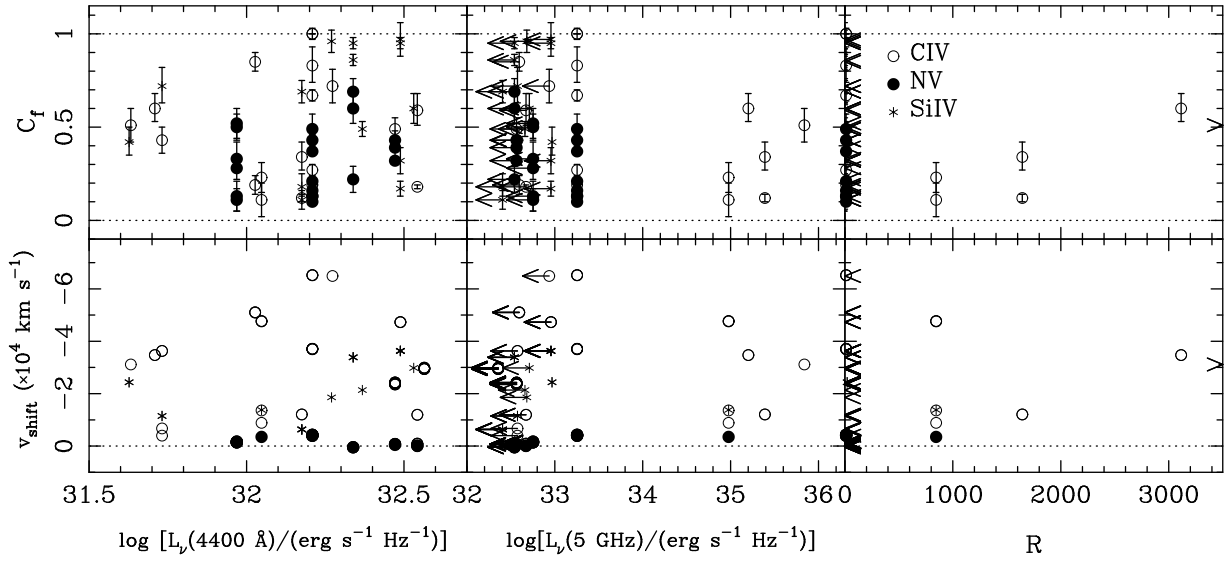}
}
\caption{Plots of NAL parameters (i.e., \cf\ and \voff) against quasar
  properties (i.e., $L_{\nu}$(4400~\AA), $L_{\nu}$(5~GHz), and
  $\rl$). Only intrinsic NAL components (class A+B), whose \cf\ values
  are physical (i.e., $0 < \cf$ and $\cf < 1$) and evaluated with high
  reliability (i.e., $\sigma$(\cf) $<$ 0.1), are plotted. For quasars
  that were not detected in the radio, we plot upper limits on
  $L_{\nu}$(5~GHz) and $\rl$. Open circles denote \ion{C}{4} NALs,
  filled circles denote \ion{N}{5} NALs, and asterisks denote
  \ion{Si}{4} NALs.\label{fig:correl}}
\end{figure}

With our large sample of intrinsic NALs we can search for relations
between the NAL properties (e.g., \cf\ $vs$ \voff\ and rest equivalent
width $vs$ \voff) and the properties of the quasars hosting them
($L_{opt}$, $L_{radio}$, $\rl$). Figure~\ref{fig:correl} shows plots
of these parameters against each other, including class A$+$B
NALs. For some quasars, we plot just the upper limits on $L_{radio}$,
since the corresponding radio sources have not been detected. We do
not find any significant correlations in these plots.  We also
searched for a correlation between the offset velocity and the
coverage fractions and rest equivalent width of \ion{C}{4}, \ion{N}{5}
\ion{Si}{4} NALs but we did not find any (see Figure~\ref{fig:cfvoff})

\subsection{Ionization State\label{sec:ionization}}

Traditionally, BALs have been classified according to their ionization
level into HiBALs (high-ionization BALs), LoBALs (low-ionization
BALs), and FeLoBALs (extremely low-ionization BALs that show
\ion{Fe}{2} lines; e.g., Weymann et al. 1991).  Motivated by the
classification of BALs and by the work of Bergeron et al. (1994) on
classifying intervening NALs, we attempt to classify intrinsic NALs
(classes A plus B) in a similar manner.  We consider the range of
conditions in the systems presented in this paper in conjunction with
our concurrent study of lower redshift intrinsic NALs in the {\it
HST}/STIS Echelle archive (Ganguly et al. 2007; in preparation). This
low redshift sample adds an important element because the \ion{O}{6}
doublet can often be examined, unlike in the spectra of $z=2$--4
quasars of this paper for which the \lya\ forest is quite thick.

We find that we can classify the intrinsic NALs in our sample into 2
categories according to the relative absorption strengths in various
transitions:

\begin{description}

\item
[\it Strong \ion{C}{4} NALs. ---] In these systems \lya\ is often
saturated and black.  \ion{N}{5} is detected in some of these
\ion{C}{4} systems, but it is weaker than \ion{C}{4}.

\item
[\it Strong \ion{N}{5} NALs. ---] The corresponding \lya\ lines have
an equivalent width that is less than twice the
\ion{N}{5}~$\lambda$1239 equivalent width.  \ion{C}{4} is detected in
some of these \ion{N}{5} systems, but is typically weaker than
\ion{N}{5}.

\end{description}

To clarify, in the case of a NAL system which has both \ion{C}{4} and
\ion{N}{5} transitions, we classify it as a strong \ion{C}{4} system
if \lya\ in the system is black or if the total equivalent width of
\lya\ is larger than twice the \ion{N}{5} equivalent width. Otherwise,
we classify them as a strong \ion{N}{5} system, even if \ion{C}{4} is
stronger than \ion{N}{5}.

One possible explanation for the physical difference between
\ion{C}{4} and \ion{N}{5} systems is a difference in ionization
parameter, however the column density, the metallicity, and/or the
metal abundance pattern could also affect these
conditions. Photoionization models will be required for the conclusive
discussion (Wu et al. 2007; in preparation).

We applied the above criteria to 28 class A and 11 class B NALs, and
classified them into 28 strong-\ion{C}{4} systems (17 class A and 11
class B) and 11 strong-\ion{N}{5} systems (all of them are class A).
The $\zabs=2.7164$ system toward HS1700$+$1055 has an \ion{O}{6}
doublet which is much stronger than the \ion{N}{5}, \ion{C}{4}, and
\lya\ lines. Although we classified it as an \ion{N}{5} intrinsic NAL,
it would likely fall in the additional category of a strong \ion{O}{6}
system, if we allowed for that category.  Finally, we added a
subclassification based on the detection of low-ionization (${\rm IP}
< 25$~eV) or intermediate-ionization (${\rm IP}=35$--50~eV) absorption
lines in the same system.  We note that BAL systems sometimes have
extremely low-ionization transitions such as \ion{Mg}{2} (${\rm
IP}=15$~eV; e.g., Hall et al. 2002).  The results are summarized in
Table~\ref{tab:ion} where we report the transitions detected in each
of the above systems. Notes on individual systems and velocity plots
are included in the Appendix.

Table~\ref{tab:ion} shows that among the 28 \ion{C}{4} intrinsic
systems, 15 have low-ionization lines, while 25 have
intermediate-ionization lines.  Two systems (at $\zabs = 2.7699$ in
Q1425$+$6039 and at $\zabs = 2.3198$ in Q1548$+$0917) are sub-DLA
systems, which casts some doubt on their intrinsic classification (see
the Appendix for a discussion of these individual cases). Nonetheless,
we see that more than half of the intrinsic \ion{C}{4} NALs have
related low ionization absorption lines.  Among the 11 \ion{N}{5}
systems, 5 have intermediate ionization absorption, and only 1 has low
ionization absorption. This is another noteworthy difference between
\ion{C}{4} and \ion{N}{5} intrinsic NALs.

In 41\%\ of intrinsic systems (16 out of 39) we detect low-ionization
transitions, and 55\% of quasars (11 out of 20) that have at least one
intrinsic system, have at least one system with low-ionization
transitions. This is much higher than the fraction of LoBAL quasars
among BAL quasars (17\% according to Sprayberry \& Foltz 1992, 13\%
according to Reichard et al. 2003).

Unfortunately, \ion{O}{6} is in the \lya\ forest or is not covered in
our spectra, making it quite difficult to establish the ionization
properties of our systems. However, an additional class of \ion{O}{6}
associated NALs (possibly intrinsic) has emerged from our study of
lower redshift quasars using the STIS/Echelle archive.  These
\ion{O}{6} systems are sometimes redward of the emission redshift of
the quasar, they have relatively weak \lya\ lines compared to
\ion{O}{6}, and they do not always have partial coverage (Ganguly et
al. 2006; in preparation).  A few of our \ion{N}{5} intrinsic NALs
with non-black \lya\ may have very strong \ion{O}{6} absorption, and
thus could fall in this third category.

\section{Discussion\label{sec:discussion}}

\subsection{Evolution of the Frequency of Intrinsic NALs With Redshift}

If we associate intrinsic NALs with quasar outflows, their evolution
with redshift traces the evolution of quasar outflows. Moreover, since
outflow properties depend on the properties of the accretion flow (at
least in disk wind models), the redshift evolution of NALs provides
constraints on the evolution of quasar fueling rates.

We begin by considering quasars with {\it strong} \ion{C}{4} AALs
($W_{rest} \geq 1.5$~\AA), which have a high probability of being
intrinsic because of their small velocity offsets from the quasars. At
intermediate redshifts ($z=1$--2), F86 found strong \ion{C}{4} AALs in
33\% of quasars, while G01 detected such a strong \ion{C}{4} AAL in
only 2\% of low-redshift quasars ($z < 1$), and suggested that this
fraction increases with redshift. In our high redshift sample
($z=2$--4), we only detect one rather strong AAL with $W_{rest} \sim
1.0$~\AA.  This could mean that the frequency of strong AALs peaks at
$z\sim 2$, which coincides with the redshift at which the quasar
density peaks. There are, however, a number of noteworthy caveats to
this apparent evolutionary trend, namely (a) the samples used at
different redshifts have different optical and radio luminosity
distributions (the G01 sample includes a significant fraction of
low-luminosity objects, for example), (b) strong AALs appear to prefer
steep-spectrum radio-loud quasars (F86; Anderson et al. 1987) which
make up different fractions of the above samples (the F86 sample is
predominantly radio-loud), and (c) strong AALs need not always be
intrinsic, in particular they could come from galaxies in a cluster
surrounding the quasar, which are more often found around radio-loud
quasars (e.g., Yee \& Green 1987, Ellison, Yee, \& Green 1991; Yee \&
Ellison 1993; Wold et al. 2000).

We now turn our attention to the evolution of the fraction of quasars
with {\it intrinsic} AALs, including weak ones, down to a rest-frame
equivalent width of $\sim 50$~m\AA\ (using samples constructed on the
basis of partial coverage or variability). At $z < 1.5$, Wise et
al. (2004) found that a minimum of 27\% of quasars have intrinsic
AALs, while at $z \sim 2$, Narayanan et al. (2004) found that a
minimum of 25\% of quasars host intrinsic AALs. Similarly, G99 find
that at least 50\% of $z \sim 2$ quasars have intrinsic AALs, and in
this study, we find that a minimum of 23\% of quasars at $z=2$--4 have
intrinsic AALs. These fractions are consistent with each other,
especially considering that some of the samples are rather small (only
six quasars in the G99 sample). An important caveat here is that all
of these results refer to lower limits since NALs that do not show
variability or partial coverage can still be intrinsic. This caveat
not withstanding, the fraction of quasars hosting {\it intrinsic} AALs
does not show significant redshift evolution.  A closely related
question is the redshift evolution of the fraction of AALs that are
intrinsic.  At $z< 1.5$ this fraction is $>21$\% (Wise et al. 2004),
at $z\sim 2$ it is $>23$\% (Narayanan et al. 2004) or $>60$\% (G99),
and at $z=2$--4 it is $>27$\% (this work). Thus, this quantity does
not appear to evolve with redshift either, but this conclusion is also
subject to the caveat noted above.

\subsection{Properties of Intrinsic Absorbers in Our Sample\label{sec:prop}}

\subsubsection{Range of Coverage Fractions}

In non-associated regions (i.e., $|\voff|> 5,000~\kms$), we found that
10--17\%\ of \ion{C}{4} NALs are intrinsic to the quasars.  Richards
et al. (1999) and Richards (2001) suggested that this fraction is as
large as 36\%, based on the excess of absorbers at $>5000~\kms$ in
steep spectrum as compared to flat spectrum radio-loud quasars. Our
estimate is strictly only a lower limit, because not all intrinsic
NALs can be identified using partial coverage.  If we take both our
result that 10--17\% of \ion{C}{4} NALs are intrinsic and the Richards
et al. number of 36\% at face value and try to reconcile them, we
conclude that only 30--50\%\ of intrinsic \ion{C}{4} NALs exhibit
detectable partial coverage in our sample.  We caution that the two
samples were selected in different ways, leading to biases.  For
example, the sample of Richards et al. (1999) has a larger fraction
($\sim 50$\%) of radio-loud quasars than our sample ($32$\%).
However, since they found that intrinsic NALs prefer radio-quiet
quasars, the difference between our estimate of the fraction of
quasars hosting intrinsic NALs and theirs would be larger than what it
appears, leading to a larger number of fully covered intrinsic NALs.

This conclusion is bolstered by the fact that the distribution of
measured coverage fractions of intrinsic \ion{C}{4} NALs spans the
range from 0 to nearly 1 (within uncertainties).  The covering factors
are presumably related to the physical sizes of the absorbing
structures, which could span a significant range.  In this sense there
is nothing special about the size corresponding to a covering factor
of 1.  If there is a cutoff in size, there is no reason it would occur
particularly at this value.  Thus we strongly expect that many
intrinsic absorbers with unit coverage fraction exist.

\subsubsection{Ionization Conditions}

The wide range of absorber ionization conditions presented in
\S\ref{sec:ionization} (e.g., ``strong \ion{N}{5}'' vs ``strong
\ion{C}{4}'' absorbers) may be a result of a wide distribution of
distances of the gas parcels from the continuum source and/or a wide
distribution of their densities. The velocity offset of a system can
serve as an indicator of its proximity to the continuum source under
the assumption that parcels of gas in an outflowing wind are
accelerated outwards from the continuum source. Detailed
photoionization modeling of selected systems can help us determine
whether there is a relation between their ionization state of NAL
systems and their apparent outflow velocity, which will allow us to
determine whether there is ionization stratification in the outflow.
The same models may also provide constraints on the density of the
gas. In fact, photoionization models for the associated absorber in
QSO J2233$-$606 (Gabel, Arav \& Kim 2006) show that the
higher-ionization kinematic components are at lower blueshifted
velocities than lower-ionization components. Moreover, preliminary
results of photoionization models for three of the quasars in our
sample (Wu et al. 2007, in preparation) suggest that absorbers at
different velocities in the same quasar can have a wide range of
densities. An alternative way of constraining the density is through
monitoring observations, which can set limits on the recombination
time scale of the absorber, hence constrain the density (see, for
example, the discussion in Wise et al. 2004, Narayanan et al. 2004,
and M05).

It is also important to ask whether the low ionization transitions in
intrinsic \ion{C}{4} NALs are aligned with the high ionization
transitions.  An inspection of the velocity plots of individual
systems in the Appendix shows that the line centers and profiles of
low-ionization lines in intrinsic NAL systems are almost always
similar to those of high-ionization lines.  This leads us to conclude
that the low-ionization lines arise in the intrinsic absorbers and not
in the ISM of the host galaxy.

\subsubsection{Intrinsic NALs at Large Velocity Offsets}

Our finding (supported by the independent method of Richards 2001)
that a significant fraction of NALs, even at high velocity offsets
from the quasar redshift, are intrinsic (10--17\%\ of \ion{C}{4} NALs
and 15--20\% of \ion{Si}{4} NALs) has implications for cosmology since
it affects the estimates of the density of intervening systems per
unit redshift ($dN/dz$).  For example, at $z=2$--4, $dN/dz$ is
estimated to be 3.11 for \ion{C}{4} and 1.36 for \ion{Si}{4} systems
with a rest frame equivalent width greater than 0.15\AA\ (Misawa et
al. 2002). If these values are reduced by 15--30\% after correction
for intrinsic systems, this would require some adjustment of models
for cosmological evolution of these systems.  Furthermore, the
contamination of the \lya\ forest by very high metallicity intrinsic
NALs could seriously bias estimations of the metallicity of the
forest.

\subsection{Implications for Models of Quasar Outflows\label{sec:implications}}

The results of our survey have direct implications for models of
quasar outflows, whatever the acceleration mechanism.

\begin{enumerate}

\item
The minimum fraction of quasars with at least one intrinsic NAL
(\ion{C}{4}, \ion{N}{5}, or \ion{Si}{4}) is 43--54\%.
If there is an additional family of intrinsic \ion{O}{6} absorbers,
without corresponding \ion{C}{4} or \ion{N}{5} lines, the minimum
fraction of quasars with intrinsic NALs will increase further. This
leads to a constraint on the solid angle subtended by the NAL
absorbers to the background source(s). We can interpret this result in
the context of a simple (but clearly not unique) picture in which all
quasars in our sample are similar to each other and the NAL gas is
located at intermediate latitudes above the streamlines of a fast,
equatorial outflow (a BAL wind). This geometry is similar to that
depicted in Figure~13 of G01, although it needs not apply specifically
to an outflow launched from the accretion disk; a pressure-driven
outflow launched from a distance of an order of 1~pc from the black
hole may have a similar geometry. In this picture, the above
observational constraint translates directly into the opening angle of
the NAL zone, given the opening angle of the BAL wind. For an opening
angle of the BAL wind of approximately 7--12$^\circ$ (10--20\% of
quasars are BAL quasars; e.g., Hamann, Korista, \& Morris 1993) the
angular width of the NAL zone turns out to be 25--27$^\circ$. In this
picture the polar region, at latitudes above the NAL zone may be
filled with highly-ionized gas. This idea is consistent with the
finding of Barthel, Tytler \& Vestergaard (1997) that strong
associated \ion{C}{4} NALs with $W_{rest}$ $>$ 3 \AA\ are most common
at angles far from the jet axis in radio-loud QSOs (as high as
$45^{\circ}$). Such strong NALs may represent intermediate viewing
angles between intrinsic NALs and BALs.  Other simple interpretations
are also possible: for example, the NAL gas may be distributed
isotropically around the central engine, or some quasars may be more
likely than others to host such absorbers (e.g., the width of the NAL
zone may depend on luminosity, as suggested by Elvis 2000).

\item
The density of intrinsic (class A$+$B) \ion{C}{4} NALs per unit
velocity interval, $dN/d\beta$, increases from 4.1 at $\voff <
-5,000~\kms$ to 18 at $\voff > -5,000~\kms$.  This is a significant
increase with a chance probability of only 5\%.
These results have a number of implications for outflow models. First,
the high-velocity NALs (71\% of class A+B NALs at $\voff <
-5,000~\kms$) are unlikely to be associated with pressure-driven
outflows because these outflows can only reach velocities of order
$10^3~\kms$. The distribution of $dN/d\beta$ with $\beta$ does not
provide a straightforward constraint on the acceleration mechanism
because the models do not predict how individual parcels of NAL gas
are distributed. One could take the higher value of $dN/d\beta$ at low
velocities as an indication of the existence of a separate population
of absorbers, which could perhaps be identified with a pressure driven
wind. On the other hand, the velocity of a magnetocentrifugal or
line-driven accretion disk wind increases smoothly with radius, which
may lead to higher density of NALs per unit velocity at low velocities
(assuming that the gas parcels are condensations that form near the
base of the wind and are accelerated outwards). Economy of means and
Occam's razor favor the latter explanation since it applies to both
high- and low-velocity NALs and invokes a single acceleration
mechanism. This issue can be addressed observationally by constraining
the distance of the low-velocity ({\it associated}) NAL gas from the
central engine. This can be done by combining constraints on its
ionization parameter (obtained from photionization models) with
constraints on its density (obtained from variability studies).

\item
The \ion{C}{4} column densities derived from fitting the line
profiles, can yield a preliminary estimate of the total hydrogen
column densities (\ion{H}{1}+\ion{H}{2}) in the absorbers. More
specifically, we can apply the calculations of Hamann (1997) to
estimate an upper limit to the hydrogen column density of absorbers in
our sample from our highest measured \ion{C}{4} column density of
$\log\; N_{\rm C\;IV} = 15.86$ (in the $z_{abs}=2.5597$ NAL in
HE0130$-$4021).  If we assume an abundance of \ion{C}{4} relative to C
of $\log\;[N_{\rm C\; IV}/N_{\rm C}] \approx -0.34$ (this limit
corresponds to the optimal ionization parameter that maximizes the
\ion{C}{4} abundance; see Figure 2 of Hamann 1997) and a solar
abundance of C relative to H, $\log\; ({\rm C}/{\rm H})=-3.44$
(Grevese \& Anders 1989), we obtain $N_{\rm H} \sim 4\times
10^{19}$~\cmm. If the C abundance is super-solar (e.g., Hamann 1997),
this limit will become smaller, of course. Higher total column
densities have also been observed in NALs; for example, Arav et
al. (2001a) find $\log N_{\rm H} > 20.3$ in
QSO~J2359$-$1241. Moreover, total column densities in BALs can be an
order of magnitude higher (e.g., in PG~0946$+$301; Arav et al.
2001b).

We emphasize that the column densities estimated above refer only to
the gas responsible for the \ion{C}{4} NALs. This gas could well have
the form of filaments embedded in a hotter medium, as predicted by
many models (see \S\ref{sec:intro} and references therein).  If that
is the case, then the total column density of the outflow can be
considerably larger than what we have estimated above.  In fact, X-ray
spectroscopy of nearby AGNs does indicate a multi-phase absorber
structure in which the column density of the hot medium is higher than
that of the cold medium by a factor of 30 or more.  (e.g., Netzer et
al. 2003, Kaspi et al. 2004, and references therein)

\end{enumerate}

For the specific case of radiatively-accelerated outflows, we can ask
whether the observed high velocities of some NALs are attainable in
the context of the model.
Hamann (1998) derives the following expression for the terminal
velocity of a radiatively accelerated wind
\begin{equation}
v_{\infty}=9,300\; r_{0.1}^{-1/2}
\left({f_{0.1} L_{46}\over N_{22}}-0.1\; M_8\right)^{1/2} \kms\; ,
\label{eqn:fred}
\end{equation}
where $r=0.1\;r_{0.1}$~pc is the distance of the absorber from the
continuum source, $L=10^{46}\; L_{46}~\ergs$ is the quasar luminosity,
$M_{BH}=10^8\; M_8\; {\rm M}_{\odot}$ is the black hole mass, and
$f=0.1\; f_{0.1}$ is the fraction of continuum photons absorbed or
scattered by the gas, and $N_H=10^{22}\; N_{22}~{\rm cm}^{-2}$ is the
column density of the absorbing gas. Noting that $L_{46}/M_8\approx
0.77\; (L/L_{Edd})$ (where $L_{Edd}$ is the Eddington luminosity) and
that $r_{0.1}/M_8\approx 2\; \xi_4$ ($\xi_4\equiv r/10^4\; r_g$ and
$r_g\equiv GM_{BH}/c^2$ is the gravitational radius), we may recast
equation~(\ref{eqn:fred}) as
\begin{equation}
v_{\infty}=5,000\; \xi_4^{-1/2}
\left({f_{0.1}\over N_{22}}\;{L\over L_{Edd}}
-0.13\right)^{1/2} \kms\;.
\label{eqn:vinf}
\end{equation}
As Hamann (1998) argues, $f_{0.1}\sim 1$ for BAL quasars. The
radiation-pressure dominated part of the accretion disk extends up to
$\xi_4\approx 0.25$ for parameters scaled as above (see Shakura \&
Sunyaev 1973). According to most models for accretion disk winds
(Murray et al 1995; Proga et al. 2000; Everett 2005), the inner launch
radius of the wind corresponds to $\xi_4\approx 0.06$--0.25.
Moreover, the \ion{C}{4} column densities that we find observationally
(see discussion above) indicate total hydrogen column densities in
NALs corresponding to $N_{22} \sim 0.01$. Under these conditions,
radiation pressure can accelerate the NAL gas to terminal speeds well
in excess of $10^5~\kms$, and the observed blueshifts of intrinsic
NALs can be easily explained.

There is an important caveat to the above estimate. We have assumed
that the column density that we have measured from the \ion{C}{4} NAL
profiles traces all of the matter that is being accelerated. But, as
we note earlier in this section, the NAL gas may be embedded in a hot
medium and it may represent only a small fraction of the total gas
mass. In such a case, the attainable terminal velocity will decrease
accordingly.  More specifically, if the NAL gas represents up to 10\%
of the total column density, then equation~(\ref{eqn:vinf}) can still
reproduce the observed NAL velocities. But if the total column density
is more than an order of magnitude higher than the \ion{C}{4} NAL
column density, then a different acceleration mechanism must be
sought.

\section{Summary and Conclusions\label{sec:summary}}

We have constructed a large, relatively unbiased, equivalent width
limited sample of {\it intrinsic} narrow absorption line (NAL) systems
found in the spectra of $z=2$--4 quasars. This sample comprises 124
\ion{C}{4}, 12 \ion{N}{5}, and 50 \ion{Si}{4} doublets, which were
separated from intervening NALs on the basis of their partial coverage
signature.  After assessing the reliability of the determination of
the intrinsic nature of these NAL systems, 28 are deemed reliably
intrinsic (``class A''), 11 are deemed possibly intrinsic (``class
B'') and 111 are deemed intervening (or unreliable intrinsic
candidates; ``class C''). Using this sample of NAL systems, we study
their demographics, the distribution of their physical properties, and
any relations between them.
We also consider the implications of these results for models of
outflowing winds. Our findings and conclusions are as follows:

\begin{enumerate}

\item
The fraction of intrinsic \ion{C}{4} systems in our sample NALs is
11--19\%.  This value increases to 33\%\ if only associated systems,
within 5,000~\kms\ of the quasar emission redshift, are
considered. This is roughly consistent with previous studies of
associated \ion{C}{4} NALs at different redshifts and employing
different methods. It is important to note, however, that all of the
above fractions are, in fact lower limits to the true fractions.  The
fraction of intrinsic \ion{Si}{4} systems is 14--18\%, although all of
these are found in the non-associated regions of the spectra, at
offset velocities $\voff < -5,000~\kms$ relative to the quasar
redshifts.  Severe contamination by the \lya\ forest allowed us to
search only for associated \ion{N}{5} systems. We found 75\%\ of such
systems to be reliably intrinsic.

\item
The minimum fraction of quasars that have one or more intrinsic NALs
is 43--54\% . While \ion{C}{4} intrinsic NALs were detected in
24--32\%\ of the quasars with adequate velocity coverage, \ion{N}{5}
intrinsic NALs were detected in 19\%\ of the quasars, and \ion{Si}{4}
intrinsic NALs in 19--24\%\ of the quasars. These are lower limits
because our spectra do not have full offset velocity coverage for
these transitions and because some intrinsic absorbers may not exhibit
the signature of partial coverage.  This places a constraint on the
solid angle subtended by the absorber to the background source(s).

\item
We find that 10--17\% of {\it non-associated} \ion{C}{4} NAL systems
are intrinsic. In cosmological applications, non-associated systems
are typically taken to be intervening. Thus, our result is important
because it shows that it is necessary to correct derived cosmological
quantities for a contamination by intrinsic NALs. A similar conclusion
was reached by Richards et al. (1999) and Richards (2001) based on a
statistical study of NALs in different quasar samples, although they
estimated a somewhat higher contamination of $\sim$36\%. Taking the
two estimates at face value (i.e., neglecting the possibility that
they differ because of systematic effects and demanding that they
should be reconciled) we are led to the conclusion that only 30--50\%\
of intrinsic \ion{C}{4} NALs exhibit partial coverage.

\item
The coverage fractions of intrinsic NALs in our sample, span almost
the entire range from 0 to 1.  Since there is a range of sizes for
intrinsic gas parcels, this also implies that there are a significant
number of intrinsic absorbers that have full coverage, and thus were
not detected in our survey.  There is no apparent relationship between
coverage fraction and velocity offset. Nor is there any relation
between the NAL properties or frequency of incidence and the
properties of the host quasars (such as optical or radio luminosity).

\item
We consider the ionization structure of the 39 class A and B intrinsic
NAL systems in our sample, and find two major categories, which may
represent absorbers of different densities and/or at different
distances from the source of the ionizing continuum.

(a) ``Strong \ion{C}{4}'' systems are characterized by strong,
partially covered \ion{C}{4} doublets, strong, usually ``black'',
\lya\ lines, and relatively weak or undetectable \ion{N}{5}
doublets. Of the 28 systems in this category, 25 also have
intermediate ionization lines detected (such as \ion{Si}{3} ,
\ion{C}{3} , and/or \ion{Si}{4} ) and 15 systems also have low
ionization absorption detected (e.g., \ion{O}{1}, \ion{Si}{2},
\ion{Al}{2}). In cases where the \lya\ profile is black, it is clear
that it does not arise in the same gas parcel as the \ion{C}{4}
absorption, since the coverage fraction for \lya\ is clearly 1. These
\ion{C}{4} systems cannot be distinguished from typical intervening
systems except by partial coverage.

(b) ``Strong \ion{N}{5}'' systems are characterized by strong
\ion{N}{5} lines, and relatively weak, non-black \lya\ lines (with
less than twice the equivalent width of their \ion{N}{5}
lines). \ion{C}{4} and \ion{O}{6} lines may also be detected in these
systems, and in some cases \ion{O}{6} may be stronger than
\ion{N}{5}. We find 11 systems in this way, 5 of which have
intermediate ionization transitions detected, and only 1 of which has
low ionization transitions detected.

\item
About 53\%\ of class A or B NAL systems include low ionization
lines. This fraction is much higher than that of LoBAL quasars among
all BAL quasars (13--17\%).  In the 15 \ion{C}{4} systems with
detected low ionization lines, the line profiles of the \ion{C}{4} and
low ionization lines are similar. In particular, low ionization lines
are rarely detected at velocities other than those of the partially
covered \ion{C}{4} components, implying that both families of lines
arise in the same parcels of gas.

\item
Our detection of a significant population of intrinsic, high-velocity
NALs ($\voff\sim 10^4\kms$) disfavors scenarios in which the absorbing
gas is associated with a pressure driven wind that does not originate
very deep in the potential well of the black hole.  The low-velocity
NALs that we have detected could still originate in pressure driven
winds. However, economy of means leads us to prefer accretion disk
winds because these can explain both the high- and the low-velocity
NALs.

\end{enumerate}

Repeated observations of the NALs in this sample at the same spectral
resolution and S/N would be particularly valuable for the following
reasons. First, they will allow us to search for NAL variability,
which can be used to confirm intrinsic NALs and to probe the relation
between variability and partial coverage of intrinsic NALs. Second,
variability sampled via multi-epoch observations carries additional
information that may allow us to constrain properties of the absorber
(e.g., density and distance from the continuum source; see
applications by Hamann et al. 1997a; Narayanan et al. 2004;
M05). Multi-epoch observations of such a large sample of NALs have
never been carried out, in spite of their utility in providing crucial
constraints on quasar outflows. The sample of NALs presented here is
an ideal starting point for such work.

\acknowledgments This work was supported by NASA grant NAG5-10817. We
would like to thank Christopher Churchill for providing us with the
{\sc minfit} software package.  UCSD work was supported in part by
NASA grant NAG5-13113 and NSF grant AST 0507717. We also thank the
anonymous referee for very useful comments and suggestions.




\clearpage

\begin{deluxetable}{llcccclcccrrr}
\rotate
\tabletypesize{\scriptsize}
\tablecaption{Sample Quasars and Their Properties\label{tab:quasars}}
\tablewidth{0pt}
\tablehead{
\colhead{(1)} & 
\colhead{(2)} & 
\colhead{(3)} &
\colhead{(4)} &
\colhead{(5)} & 
\colhead{(6)} & 
\colhead{(7)} &
\colhead{(8)} &
\colhead{(9)} &
\colhead{(10)} &
\colhead{(11)} &
\colhead{(12)} &
\colhead{(13)} \\
\colhead{QSO$^{a}$} & 
\colhead{$z_{em}$} & 
\colhead{$m_V^{b}$} &
\colhead{$m_R^{c}$} & 
\colhead{$f_{\nu}({\rm 4400\;\AA})$\tablenotemark{d}} &
\colhead{$f_{\nu}({\rm radio})$\tablenotemark{e}} &
\colhead{$\nu$$^{f}$} &
\colhead{$f_{\nu}({\rm 5\;GHz})$\tablenotemark{g}} & 
\colhead{$\rl$\tablenotemark{h}} & 
\colhead{L/Q\tablenotemark{i}} & 
\colhead{$v_{up}$(\ion{N}{5})\tablenotemark{j}} &
\colhead{$v_{up}$(\ion{Si}{4})\tablenotemark{j}} &
\colhead{$v_{up}$(\ion{C}{4})\tablenotemark{j}} \\
\colhead{} & 
\colhead{} &
\colhead{(mag)} & 
\colhead{(mag)} &
\colhead{(mJy)} &
\colhead{(mJy)} & 
\colhead{(GHz)} & 
\colhead{(mJy)} & 
\colhead{} & 
\colhead{} &
\colhead{(\kms)} &
\colhead{(\kms)} &
\colhead{(\kms)} \\
}
\startdata
Q0004$+$1711  & 2.890  & 18.70 &      & 0.051 & 159     & 4.85 &     104   &   2032   & L &   12,847 & $-$22,455 & $-$53,447 \\
Q0014$+$8118  & 3.387  &       & 16.1 & 0.394 & 692.8   & 1.4  &     182.4 &   463    & L &   33,519 &  $-$1,693 & $-$33,084 \\
Q0054$-$2824  & 3.616  &       & 17.8 & 0.080 & $<$ 2.5 & 1.4  & $<$ 0.65  & $<$ 8.11 & Q &   38,679 &     3,543 & $-$27,903 \\
HE0130$-$4021 & 3.030  & 17.02 &      & 0.235 & 4       & 5.0  &   2.6     &   11.2   & Q &   57,894 &    23,231 &  $-$8,247 \\
Q0241$-$0146  & 4.040  & 18.20 &      & 0.137 & $<$ 2.5 & 1.4  & $<$ 0.63  & $<$ 4.61 & Q &   29,887 &  $-$5,365 & $-$36,707 \\
HE0322$-$3213 & 3.302  & 17.80 &      & 0.110 & $<$ 2.5 & 1.4  & $<$ 0.66  & $<$ 6.01 & Q &    1,157 & $-$34,048 & $-$64,690 \\
Q0336$-$0143  & 3.197  &       & 18.8 & 0.034 &	446     & 4.85 &   284     &    8460  & L &   60,999 &    26,440 &  $-$5,017 \\
Q0450$-$1310  & 2.300  & 16.50 &      & 0.424 & $<$ 2.5 & 1.4  & $<$ 0.72  & $<$ 1.69 & Q &   54,351 &    19,577 & $-$11,915 \\
Q0636$+$6801  & 3.178  &       & 16.9 & 0.194 &	499     & 4.85 &   318     &   1643   & L &   68,059 &    33,768 &     2,386 \\
Q0642$+$4454  & 3.408  &       & 18.4 & 0.047 &	1204    & 4.85 &   755.3   &   16003  & L &   46,303 &    11,318 & $-$20,172 \\
Q0805$+$0441  & 2.880  & 18.16 &      & 0.084 & 401     & 4.85 &   261     &   3115   & L &   74,304 &    40,285 &     9,000 \\
Q0831$+$1248  & 2.734  & 18.10 &      & 0.091 & $<$ 2.5 & 1.4  & $<$ 0.69  & $<$ 7.62 & Q &   84,998 &    51,520 &    20,478 \\
HE0940$-$1050 & 3.080  & 16.90 &      & 0.260 & $<$ 2.5 & 1.4  & $<$ 0.67  & $<$ 2.58 & Q &   52,404 &    17,574 & $-$13,922 \\
Q1009$+$2956  & 2.644  & 16.40 &      & 0.440 & $<$ 2.5 & 1.4  & $<$ 0.70  & $<$ 1.58 & Q &    6,945 & $-$28,322 & $-$59,148 \\
Q1017$+$1055  & 3.156  &       & 17.2 & 0.147 &	195     & 4.85 &   125     &    845   & L &   59,742 &    25,139 &  $-$6,327 \\
Q1055$+$4611  & 4.118  & 17.70 &      & 0.215 & $<$ 2.5 & 1.4  & $<$ 0.63  & $<$ 2.92 & Q &   25,319 &  $-$9,969 & $-$41,237 \\
HS1103$+$6416 & 2.191  & 15.42 &      & 1.168 & $<$ 2.5 & 1.4  & $<$ 0.72  & $<$ 0.62 & Q &  109,239 &    77,353 &    47,217 \\
Q1107$+$4847  & 3.000  & 16.60 &      & 0.347 & $<$ 2.5 & 1.4  & $<$ 0.68  & $<$ 1.95 & Q &   64,739 &    30,317 &  $-$1,106 \\
Q1157$+$3143  & 2.992  & 17.00 &      & 0.240 & $<$ 2.5 & 1.4  & $<$ 0.68  & $<$ 2.82 & Q &   66,236 &    31,872 &       466 \\
Q1208$+$1011\tablenotemark{k} & 3.803  &       & 17.2 & 0.040 & $<$ 2.5 & 1.4  & $<$ 0.64  & $<$ 16.0 & Q &   10,885 & $-$24,409 & $-$55,348 \\
Q1244$+$1129  & 2.960  & 17.70 &      & 0.127 & $<$ 2.5 & 1.4  & $<$ 0.68  & $<$ 5.36 & Q & $-$1,578 & $-$36,745 & $-$67,293 \\
Q1251$+$3644  & 2.988  & 19.00 &      & 0.038 & $<$ 2.5 & 1.4  & $<$ 0.68  & $<$ 17.8 & Q &   66,522 &    32,169 &       767 \\
Q1330$+$0108  & 3.510  &       &18.56 & 0.040 & $<$ 2.5 & 1.4  & $<$ 0.65  & $<$ 16.2 & Q &   42,793 &     7,733 & $-$23,742 \\
Q1334$-$0033  & 2.801  & 17.30 &      & 0.187 & $<$ 2.5 & 1.4  & $<$ 0.69  & $<$ 3.67 & Q &   79,164 &    45,379 &    14,193 \\
Q1337$+$2832  & 2.537  & 19.30 &      & 0.031 & $<$ 2.5 & 1.4  & $<$ 0.70  & $<$ 25.7 & Q &   21,637 & $-$13,668 & $-$44,865 \\
Q1422$+$2309\tablenotemark{l} & 3.611  &       & 15.3 & 0.052 &	503     & 4.85 &   20.2    &   389    & L &   23,565 & $-$11,732 & $-$42,967 \\
Q1425$+$6039  & 3.165  &       & 16.0 & 0.449 & $<$ 2.5 & 1.4  & $<$ 0.67  & $<$ 1.50 & Q &   53,082 &    18,272 & $-$13,224 \\
Q1442$+$2931  & 2.670  & 16.20 &      & 0.526 & $<$ 2.5 & 1.4  & $<$ 0.69  & $<$ 1.32 & Q &   89,302 &    56,071 &    25,152 \\
Q1526$+$6701  & 3.020  & 17.20 &      & 0.199 & 417     & 4.85 &   269     &   1350   & L &     $-$4 & $-$35,193 & $-$65 796 \\
Q1548$+$0917  & 2.749  & 18.00 &      & 0.099 & $<$ 2.5 & 1.4  & $<$ 0.69  & $<$ 6.96 & Q &   83,443 &    49,881 &    18,797 \\
Q1554$+$3749  & 2.664  & 18.19 &      & 0.084 & 6.64    & 1.4  &   1.84    &   21.9   & Q &   14,877 & $-$20,432 & $-$51,475 \\
HS1700$+$6416 & 2.722  & 16.13 &      & 0.557 & $<$ 2.5 & 1.4  & $<$ 0.69  & $<$ 1.24 & Q &   85,440 &    51,987 &    20,956 \\
Q1759$+$7539  & 3.050  & 16.50 &      & 0.378 & 145     & 5.0  &   95.3    &   252    & L &    1,953 & $-$33,262 & $-$63,931 \\
Q1937$-$1009  & 3.806  &       & 16.7 & 0.215 & 750     & 4.85 &   458     &   2130   & L &   66,152 &    31,784 &       378 \\
HS1946$+$7658 & 3.051  & 16.20 &      & 0.498 & $<$ 2.5 & 1.4  & $<$ 0.67  & $<$ 1.35 & Q &   67,075 &    32,744 &     1,349 \\
Q2223$+$2024  & 3.560  &       & 18.5 & 0.042 &	143     & 4.85 &   88.8    &   2130   & L &   42,725 &     7,663 & $-$23,811 \\
Q2344$+$1228  & 2.763  & 17.50 &      & 0.157 & $<$ 2.5 & 1.4  & $<$ 0.69  & $<$ 4.40 & Q &   17,377 & $-$17,934 & $-$49,038 \\
\enddata
\tablenotetext{a}{Quasar names are based on B1950 coordinates.}
\tablenotetext{b}{V magnitude from V\'eron-Cetty \& V\'eron (2003).}
\tablenotetext{c}{R magnitude from USNO-A2.0 Catalog (Monet et
  al. 1998), except for Q1330$+$0108, whose R magnitude comes from the
  USNO-B Catalog (Monet et al. 2003).}
\tablenotetext{d}{Optical flux density at 4400~\AA\ in the quasar rest
  frame (converted from observed flux assuming $f_{\nu}\propto
  \nu^{-0.44}$; see \S\ref{sec:obs}).  The flux densities of
  Q0241$-$0146 and Q1055$+$4611 were corrected for \lya\ forest
  contamination as described in \S\ref{sec:obs} of the text.}
\tablenotetext{e}{Observed radio flux density from NVSS (1.4GHz;
  Condon et al. 1998), Griffith et al. (1995; 4.85GHz), or Becker,
  White, \& Edwards (1991; 4.85GHz). If no radio source is detected
  within 10$^{\prime\prime}$ of the optical source, we use the
  detection limit of the survey as an upper limit to the radio flux.
  The radio fluxes of HE0130$-$4021 and Q1759$+$7539 come from Smith
  \& Wright (1980) and Hook et al. (1996), respectively.}
\tablenotetext{f}{Radio frequency corresponding to the observed radio
  flux density.}
\tablenotetext{g}{Radio flux density at 5~GHz in the quasar rest
  frame, obtained from the observed flux, assuming
  that$f_{\nu}\propto\nu^{0.7}$.}
\tablenotetext{h}{Radio loudness parameter, defined in \S\ref{sec:obs}
  of the text.}
\tablenotetext{i}{radio-loud or radio-quiet quasar (see
  \S\ref{sec:obs} for definition).}
\tablenotetext{j}{Red limit of the velocity window in which we
  searched for \ion{C}{4}, \ion{N}{5}, or \ion{Si}{4} NALs.}
\tablenotetext{k}{This lensed quasar is amplified by a factor of
  $\sim$3.1 (Barvainis \& Ivison 2002).}
\tablenotetext{l}{This lensed quasar is amplified by a factor of
  15.38 (Kormann et al. 1994).}
\end{deluxetable}


\begin{deluxetable}{ccccc}
\tablecaption{Census of Poisson Systems, NALs, and Kinematic Components\label{tab:census}}
\tablewidth{0pt}
\tablehead{
\colhead{(1)} & 
\colhead{(2)} & 
\colhead{(3)} &
\colhead{(4)} &
\colhead{(5)} \\
\colhead{Ion} & 
\colhead{Class} &
\colhead{Poisson Systems$^{a}$} &
\colhead{Lines} &
\colhead{Components$^{b}$}
}
\startdata
All   &  All$^{c}$        & 259 & 366 & 871       \\
      & Homogeneous$^{d}$ & 150 & 206 & 706 (313) \\
      &   A$^{d}$         &  28 &  30 &  85 (61)  \\
      &   B$^{d}$         &  11 &  11 &  47 (18)  \\
      &   C$^{d}$         & 111 & 165 & 574 (234) \\
\noalign{\vskip 6pt}
\hline
\noalign{\vskip 6pt}
\ion{C}{4}  &  All$^{c}$  & 224 & 261 & 611       \\
      & Homogeneous$^{d}$ & 124 & 138 & 483 (209) \\
      &   A$^{d}$         &  14 &  14 &  33  (20) \\
      &   B$^{d}$         &   9 &   9 &  42  (14) \\
      &   C$^{d}$         & 101 & 115 & 408 (175) \\
\noalign{\vskip 6pt}
\hline
\noalign{\vskip 6pt}
\ion{N}{5}   &  All$^{c}$ &  13 &  13 &  42       \\
      & Homogeneous$^{d}$ &  12 &  12 &  41  (27) \\
      &   A$^{d}$         &   9 &   9 &  31  (25) \\
      &   B$^{d}$         &   0 &   0 &   0   (0) \\
      &   C$^{d}$         &   3 &   3 &  10   (2) \\
\noalign{\vskip 6pt}
\hline
\noalign{\vskip 6pt}
\ion{Si}{4} &  All$^{c}$  &  72 &  92 & 218       \\
      & Homogeneous$^{d}$ &  50 &  56 & 182  (77) \\
      &   A$^{d}$         &   7 &   7 &  21  (16) \\
      &   B$^{d}$         &   2 &   2 &   5   (4) \\
      &   C$^{d}$         &  41 &  47 & 156  (57)
\enddata
\tablenotetext{a}{Combined multiple NALs that lie within 200~\kms\ of
  each other. See discussion in \S\ref{sec:sample} of the text.}
\tablenotetext{b}{Narrow kinematic components deblended by Voigt profile
  fitting with {\sc minfit}. Numbers in parenthesis are number of
  components whose \cf\ values are physical (i.e., $0 < \cf$ and $\cf\
  < 1$).}
\tablenotetext{c}{All lines that are detected at greater than a
  5$\sigma$ confidence level, $W_{obs}/\sigma(W_{obs}) \geq 5$,
  regardless of their equivalent width.}
\tablenotetext{d}{Doublets with rest-frame equivalent widths of their
  stronger (bluer) components larger than the minimum equivalent width
  that correspond to the 5$\sigma$ detection limit, $W_{min}$.  These
  limits are $W_{min}\;$(\ion{C}{4})$\;= 0.056$~\AA ,
  $W_{min}\;$(\ion{N}{5})$\;=0.038$~\AA , and
  $W_{min}\;$(\ion{Si}{4})$\;=0.054$~\AA\ (see \S\ref{sec:sample} of
  the text).}
\end{deluxetable}


\begin{deluxetable}{lcccccccccccc}
\tabletypesize{\footnotesize}
\setlength{\tabcolsep}{0.02in}
\tablecaption{Statistical Properties of Poisson Systems of NALs\label{tab:stats}}
\tablewidth{0pt}
\tablehead{
\colhead{(1)} & 
\colhead{(2)} & 
\colhead{(3)} &
\colhead{(4)} &
\colhead{(5)} &
\colhead{} & 
\colhead{(6)} & 
\colhead{(7)} &
\colhead{(8)} &
\colhead{} &
\colhead{(9)} & 
\colhead{(10)} & 
\colhead{(11)} \\
\multicolumn{2}{c}{} &
\multicolumn{3}{c}{\ion{C}{4}} &
\multicolumn{1}{c}{} & 
\multicolumn{3}{c}{\ion{N}{5}}  &
\multicolumn{1}{c}{} & 
\multicolumn{3}{c}{\ion{Si}{4}} \\
\noalign{\vskip 3pt}
\cline{3-5}
\cline{7-9}
\cline{11-13} 
\noalign{\vskip 3pt}
\multicolumn{2}{c}{} &
\colhead{AAL} & 
\colhead{non-AAL} &
\multicolumn{1}{c}{total} & 
\colhead{} & 
\colhead{AAL} &
\colhead{non-AAL} & 
\multicolumn{1}{c}{total} & 
\colhead{} & 
\colhead{AAL} &
\colhead{non-AAL} & 
\multicolumn{1}{c}{total} 
}
\startdata
     & $\delta z^a$&  0.64   & 17.      & 18.     & &  2.3    & 0.32   &  2.6    & & 1.3     & 12.     & 14.     \\
  & $\delta\beta^a$&  0.17   &  4.9     &  5.1    & &  0.58   & 0.079  &  0.66   & & 0.32    &  3.3    &  3.6    \\
\noalign{\vskip 6pt}
\hline
\noalign{\vskip 6pt}
           &  $N^b$&  3   (3)& 11    (8)& 14   (9)& &  9   (7)& 0   (0)&  9   (7)& & 0    (0)&  7   (7)&  7   (7)\\
class A    &$dN/dz$&  4.7    &  0.63    & 0.78    & &  3.9    & 0.0    &  3.4    & & 0.0     & 0.57    & 0.5     \\
       &$dN/d\beta$& 18.     &  2.3     & 2.8     & & 16.     & 0.0    & 14.     & & 0.0     &  2.2    & 2.0     \\
\noalign{\vskip 6pt}
\hline			  			      
\noalign{\vskip 6pt}
           &  $N^b$&  3   (3)& 20   (11)& 23  (12)& &  9   (7)& 0   (0)&  9   (7)& & 0    (0)&  9   (9)&  9   (9)\\
class A$+$B&$dN/dz$&  4.7    &  1.2     & 1.3     & &  3.9    & 0.0    &  3.4    & & 0.0     &  0.73   & 0.66    \\
       &$dN/d\beta$& 18.     &  4.1     & 4.6     & & 16.     & 0.0    & 14.     & & 0.0     &  2.8    & 2.5     \\
\noalign{\vskip 6pt}
\hline			  			      
\noalign{\vskip 6pt}
           &  $N^b$&  6   (5)& 95   (34)& 101 (34)& &  3   (3)& 0   (0)&  3   (3)& &  4   (4)& 37  (24)& 41  (24)\\
class C    &$dN/dz$&  9.4    &  5.5     &  5.6    & &  1.3    & 0.0    &  1.1    & &  3.2    &  3.0    &  3.0    \\
       &$dN/d\beta$& 36.     & 19.      & 20.     & &  5.2    & 0.0    &  4.6    & & 13.     & 11.     & 11.     \\
\noalign{\vskip 6pt}
\hline			  			      
\noalign{\vskip 6pt}
           &  $N^b$&  9  (13)& 115  (37)& 124 (37)& & 12  (37)& 0   (0)& 12  (37)& &  4  (24)& 46  (37)& 50  (37)\\
Total      &$dN/dz$& 14.     &  6.6     &  6.9    & &  5.2    & 0.0    &  4.6    & &  3.2    &  3.7    &  3.7    \\
       &$dN/d\beta$& 54.     & 24.      & 25.     & & 21.     & 0.0    & 18.     & & 13.     & 14.     & 14.     \\
\noalign{\vskip 6pt}
\hline
\noalign{\vskip 6pt}
NALs$^c$  & A/Total& 33\%    & 10\%     & 11\%    & & 75\%    & \dots  & 75\%    & & 0\%     & 15\%    & 14\%    \\
        & A+B/Total& 33\%    & 17\%     & 19\%    & & 75\%    & \dots  & 75\%    & & 0\%     & 20\%    & 18\%    \\
\noalign{\vskip 6pt}
\hline
\noalign{\vskip 6pt}
Quasars$^d$&A/Total& 23\%    & 22\%     & 24\%    & & 19\%    & \dots  & 19\%    & & 0\%     & 19\%    & 19\%    \\
        & A+B/Total& 23\%    & 30\%     & 32\%    & & 19\%    & \dots  & 19\%    & & 0\%     & 24\%    & 24\%
\enddata

\tablenotetext{a}{The total redshift and speed intervals considered in
  the determination of $dN/dz$ and $dN/d\beta$}
\tablenotetext{b}{$N$ denotes the number of Poisson systems (i.e.,
  groups of NALs that lie within 200~\kms\ of each other; see
  \S\ref{sec:sample}). The number in parenthesis gives the number of 
  quasars in which these Poisson systems are found.}
\tablenotetext{c}{Percentage of class A and class A+B NALs relative to 
  all NALs, broken down by transition and by velocity window (associated 
  $vs$ non-associated).}
\tablenotetext{d}{Percentage of quasars hosting class A and class A+B
  NALs relative to all quasars, broken down by transition and by
  velocity window (associated $vs$ non-associated).}
\end{deluxetable}


\begin{deluxetable}{lccccccl}
\tabletypesize{\small}
\tablecaption{Ionization Levels of Class A and Class B Systems$^{a}$\label{tab:ion}}
\tablewidth{0pt}
\tablehead{
\colhead{(1)} & 
\colhead{(2)} & 
\colhead{(3)} &
\colhead{(4)} &
\colhead{(5)} & 
\colhead{(6)} & 
\colhead{(7)} &
\colhead{(8)} \\
\colhead{} & 
\colhead{} &
\colhead{Low-Ion.$^b$} &
\colhead{Interm.-Ion.$^c$} & 
\colhead{\ion{C}{4}} &
\colhead{\ion{N}{5}} &
\colhead{\ion{O}{6}} &
\colhead{Ionization} \\
\colhead{QSO} & 
\colhead{\zabs} & 
\colhead{13--24 eV} & 
\colhead{33--48 eV} &
\colhead{65 eV} &
\colhead{98 eV} & 
\colhead{138 eV} & 
\colhead{Class$^d$}
}
\startdata
\multicolumn{8}{c}{Class A Systems} \\
\noalign{\vskip 3pt \hrule \vskip 3pt}
HE0130$-$4021     & 2.5597     & N        & Y          & Y        & N        & \dots    & 1, HI  \\
                  & 2.9749     & N        & N          & \dots    & Y        & Y        & 2, H   \\
HE0322$-$3213     & 3.2781     & N        & N          & \dots    & Y        & \dots    & 2, H   \\
                  & 3.2818     & \dots    & Y          & \dots    & Y        & \dots    & 2, HI  \\
Q0450$-$1310      & 2.2307     & N        & Y          & \dots    & Y        & \dots    & 2, HI  \\
Q0636$+$6801      & 3.0134     & N        & Y          & Y        & \dots    & \dots    & 1, HI  \\
Q0642$+$4454      & 2.9721     & N        & Y          & Y        & \dots    & \dots    & 1, HI  \\
Q0805$+$0441      & 2.4544     & N        & Y          & Y        & N        & \dots    & 1, HI  \\
                  & 2.6517     & Y        & Y          & Y        & Y        & \dots    & 1, HIL \\
                  & 2.8589     & N        & N          & Y        & Y        & Y        & 2, H   \\
Q1009$+$2956      & 2.2533     & Y        & Y          & \dots    & N        & \dots    & 1, HIL \\
                  & 2.6495     & N        & Y          & \dots    & Y        & Y        & 2, HI  \\
Q1017$+$1055   & $\approx$2.97 & Y$^e$    & Y$^e$      & Y$^e$    & \dots    & \dots    & 2, HIL \\
               & $\approx$3.03 & N        & N          & Y$^e$    & \dots    & \dots    & 2, H   \\
               & $\approx$3.11 & N        & Y          & \dots    & Y$^e$    & \dots    & 2, HI  \\
Q1055$+$4611      & 3.5314     & Y        & Y          & \dots    & \dots    & \dots    & 1, HIL \\
Q1107$+$4847      & 2.7243     & Y        & Y          & Y        & N        & \dots    & 1, HIL \\
Q1330$+$0108      & 3.1148     & Y        & Y          & Y        & \dots    & \dots    & 1, HIL \\
Q1425$+$6039      & 2.7699     & Y        & Y          & Y        & \dots    & \dots    & 1, HIL (DLA)\\
Q1548$+$0917      & 2.3198     & Y        & Y          & Y        & N        & \dots    & 1, HIL (DLA)\\
                  & 2.6659     & Y        & Y          & Y        & N        & \dots    & 1, HIL \\
                  & 2.6998     & Y        & Y          & Y        & N        & N        & 1, HIL \\
Q1554$+$3749      & 2.3777     & Y        & Y          & \dots    & N        & \dots    & 1, HIL \\
HS1700$+$6416     & 2.7125     & N        & N          & Y        & Y        & \dots    & 2, H   \\
                  & 2.7164     & N        & N          & N        & Y        & Y        & 2, H   \\
HS1946$+$7658     & 2.8928     & Y        & Y          & Y        & N        & \dots    & 1, HIL \\
                  & 3.0385     & N        & N          & Y        & N        & \dots    & 1, H   \\
                  & 3.0497     & Y        & Y          & Y        & Y        & \dots    & 1, HIL \\
\noalign{\vskip 3pt \hrule \vskip 3pt}
\multicolumn{8}{c}{Class B Systems} \\
\noalign{\vskip 3pt \hrule \vskip 3pt}
HE0130$-$4021     & 2.2316     & N        & Y          & Y        & \dots    & \dots    & 1, HI  \\
Q0241$-$0146      & 3.0451     & N        & Y          & Y        & \dots    & \dots    & 1, HI  \\
HE0940$-$1050     & 2.8347     & N        & Y          & Y        & \dots    & \dots    & 1, HI  \\
Q1017$+$1055      & 2.5408     & N        & N          & Y        & N        & \dots    & 1, H   \\
Q1055$+$4611      & 3.3658     & N        & N          & Y        & \dots    & \dots    & 1, H   \\
HS1103$+$6416     & 1.8874     & N        & Y          & Y        & N        & \dots    & 1, HI  \\
                  & 1.8919     & Y        & Y          & Y        & N        & \dots    & 1, HIL \\
Q1334$-$0033      & 2.2010     & N        & Y          & Y        & N        & \dots    & 1, HI  \\
Q1548$+$0917      & 2.6082     & Y        & Y          & Y        & N        & \dots    & 1, HIL \\
HS1700$+$6416     & 2.4330     & Y        & Y          & Y        & N        & \dots    & 1, HIL \\
                  & 2.4394     & N        & Y          & Y        & N        & \dots    & 1, HI  \\
\enddata
\tablenotetext{a}{A ``Y'' indicates that the system contains one or
  more lines of a given type are detected. A ``N'' indicates that no
  lines from that category were detected, while a blank entry means
  that either the spectrum did not cover those transitions or the
  transitions were severely blending with \lya\ forest or data defects.}
\tablenotetext{b}{Low-ionization (from ions with ionization potentials
  between 13 and 24~eV): \ion{O}{1}~$\lambda$1302, \ion{Si}{2}~$\lambda$1190,
  $\lambda$1193, $\lambda$1260, $\lambda$1527, \ion{Al}{2}~$\lambda$1671,
  \ion{C}{2}~$\lambda$1335. }
\tablenotetext{c}{Intermediate-ionization lines(from ions with
  ionization potentials between 33 and 48~eV):\ion{Si}{3}~$\lambda$1207,
  \ion{Si}{4}~$\lambda$1394, $\lambda$1403, \ion{C}{3}~$\lambda$977. }
\tablenotetext{d}{1=\ion{C}{4} dominated system, 2=\ion{N}{5} dominated
  system. H, I, and L mean that the system contains high-,
  intermediate-, and low-ionization lines, respectively.}
\tablenotetext{e}{These lines in Q1017$+$1055 are classified as
  ``mini-BALs'' because they have widths of 500-1,000~\kms. See
  detailed discussion in the Appendix.}
\end{deluxetable}


\clearpage

\begin{deluxetable}{lccccccccccccl}
\rotate
\tabletypesize{\scriptsize}
\setlength{\tabcolsep}{0.04in}
\tablecaption{Properties of Narrow Absorption Lines\label{tab:master}}
\tablewidth{0pt}
\tablehead{
\colhead{(1)} & 
\colhead{(2)} & 
\colhead{(3)} &
\colhead{(4)} &
\colhead{(5)} & 
\colhead{(6)} & 
\colhead{(7)} &
\colhead{(8)} &
\colhead{(9)} &
\colhead{(10)} &
\colhead{(11)} &
\colhead{(12)} &
\colhead{(13)} &
\colhead{(14)} \\
\colhead{} & 
\colhead{} &
\colhead{} &
\colhead{} &
\colhead{$\lambda_{obs}$$^b$} & 
\colhead{$W_{obs}$$^c$} & 
\colhead{$z_{abs}$$^d$} &
\colhead{$\voff^e$} &
\colhead{$\log N$} &
\colhead{$\sigma(v)$} &
\colhead{} & 
\colhead{Reliability} & 
\colhead{} & 
\colhead{Other} \\
\colhead{QSO} & 
\colhead{$z_{em}$} &
\colhead{AAL$^a$} &
\colhead{Ion} & 
\colhead{(\AA)} &
\colhead{(\AA)} &
\colhead{} &
\colhead{(\kms)} &
\colhead{(\cmm)} &
\colhead{(\kms)} &
\colhead{$C_{f}$$^f$} &
\colhead{Class$^g$} & 
\colhead{Sample$^h$} & 
\colhead{Ions$^i$} 
}
\startdata
Q0004+1711	&2.890	&       &\ion{C}{4}     &4843.0     &2.7768     &2.1281    &$-$64375     &               &82.0            &                       &C1   &L,H   &\lya (\ion{C}{2}~$\lambda$1335, \ion{Si}{3}~$\lambda$1207, \\
                &       &       &               &           &           &2.1271    &$-$64473     &13.71$\pm$0.22 &19.3$\pm$2.9    &0.49$^{+0.34}_{-0.17}$ &     &      &\ion{Si}{4})                                               \\
                &       &       &               &           &           &2.1275    &$-$64431     &14.21$\pm$0.05 &11.9$\pm$0.7    &0.97$^{+0.07}_{-0.07}$ &     &      &                                        \\
                &       &       &               &           &           &2.1280    &$-$64385     &14.48$\pm$0.04 &22.1$\pm$1.6    &0.98$^{+0.07}_{-0.07}$ &     &      &                                        \\
                &       &       &               &           &           &2.1283    &$-$64357     &12.44$\pm$1.80 &7.4$\pm$13.4    &1.00                   &     &      &                                        \\
                &       &       &               &           &           &2.1286    &$-$64334     &14.28$\pm$0.05 &15.3$\pm$3.3    &1.00                   &     &      &                                        \\
                &       &       &               &           &           &2.1289    &$-$64309     &12.74$\pm$1.32 &11.0$\pm$13.1   &1.00                   &     &      &                                        \\
                &       &       &               &           &           &2.1290    &$-$64298     &13.88$\pm$0.06 &10.8$\pm$0.7    &1.00$^{+0.08}_{-0.07}$ &     &      &                                        \\
 \cline{3-14} \\				       		   			 				    	   		    
		&	&	&\ion{Si}{4}    &4903.0     &1.0584     &2.5178    &$-$30068     &               &42.6            &                       &C1   &L,H   &\lya (\ion{Si}{2}~$\lambda\lambda$1190,1193,               \\
                &       &       &               &           &           &2.5171    &$-$30132     &12.30$\pm$0.12 &2.7$\pm$2.9     &1.00                   &     &      &\ion{Si}{2}~$\lambda$1260, \ion{Si}{3}~$\lambda$1207,      \\
                &       &       &               &           &           &2.5177    &$-$30082     &13.97$\pm$0.07 &11.5$\pm$0.5    &0.99$^{+0.09}_{-0.09}$ &     &      &\ion{Si}{4})                                               \\
                &       &       &               &           &           &2.5182    &$-$30036     &13.52$\pm$0.07 &11.4$\pm$0.6    &0.85$^{+0.09}_{-0.09}$ &     &      &                                        \\
\enddata
\tablenotetext{a}{Associated absorption line, $|\voff|\leq 5,000~\kms$.}
\tablenotetext{b}{Wavelength of flux-weighted line center.}
\tablenotetext{c}{Observed frame equivalent width of blue member.}
\tablenotetext{d}{Redshift of flux weighted line center.}
\tablenotetext{e}{Velocity offset from quasar emission redshift.}
\tablenotetext{f}{Coverage fraction.}
\tablenotetext{g}{Partial coverage reliability classes, as defined
  in \S\ref{sec:parcov}.}
\tablenotetext{h}{Subsamples: L=radio-loud, Q=radio-quiet,
  A=associated (within 5,000~\kms\ of \zem), and
  H=homogeneous/complete sample with $W_{rest}/\sigma(W_{rest} \geq 5)$.}
\tablenotetext{i}{Other lines that are detected in the system. Lines
  in parenthesis are in \lya\ forest, i.e., less reliable.}
\tablenotetext{j}{Evaluated with red member, because blue member
  blends with other lines, data defects, or echelle order gaps.}
\tablenotetext{k}{\lya, and \ion{N}{5} seem to have disappeared 
  (``normalized out'').}
\tablenotetext{l}{Voigt profile fit is not applied because of smooth
  wide absorption profile.}
\tablenotetext{m}{Approximate value.}
\tablenotetext{n}{Fitting assuming $\cf=1$.}
\end{deluxetable}


\clearpage

\clearpage

\begin{figure}
\centerline{
\includegraphics[scale=.55]{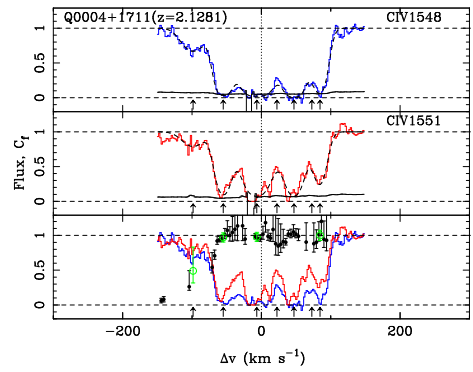}	
}
\caption{(Sample) Results of partial coverage analysis applied to 206
\ion{C}{4}, \ion{N}{5}, and \ion{Si}{4} doublets. Velocities are
defined relative to the flux-weighted center of the system, as given
in the first line of each block of Table~\ref{tab:master}. The first
two panels of each set show the profiles of the blue and red members
of a doublet on a common velocity scale, with the model profile
produced by {\sc minfit} superposed as a dashed line. The positions of
the kinematic components making up the model are marked with upward
arrows in the bottom of each panel.  The third panel of each set shows
the two profiles together, along with the resulting coverage fractions
and their 1$\sigma$ error bars.  The coverage fraction determined by
the pixel-by-pixel method is plotted in the form of black filled
circles, while the coverage fraction determined by the fitting method
is plotted in the form of green open circles.\label{fig:pcovplots}}
\end{figure}

\clearpage

\begin{figure}
\centerline{
\includegraphics[width=20cm,angle=270]{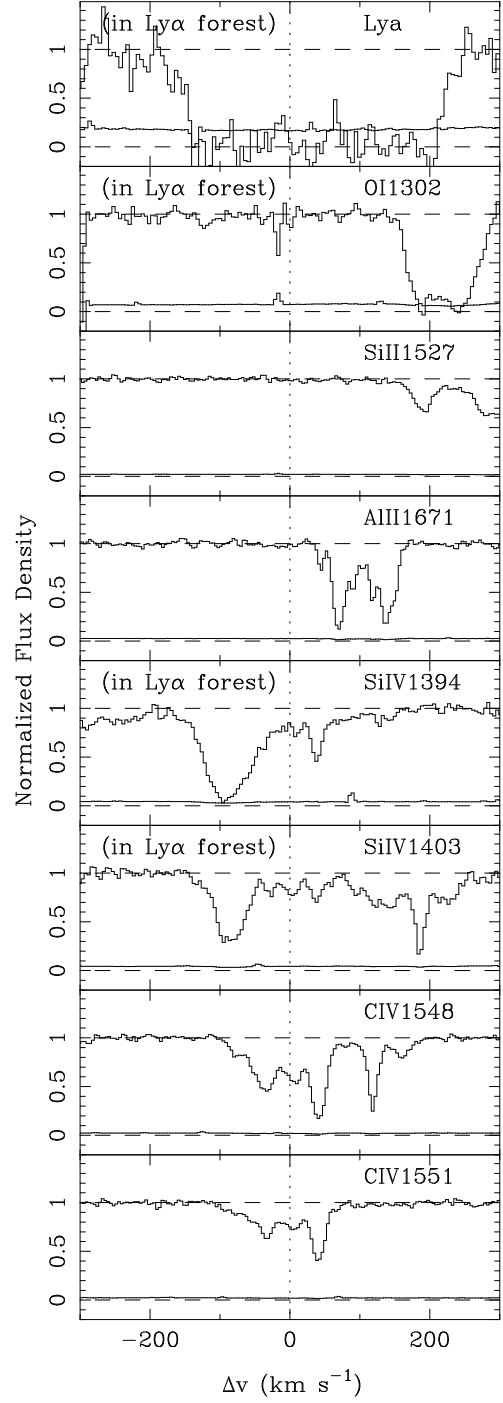}
}
\caption{(Sample) Velocity plots of various lines detected in the
class B NAL system at \zabs\ = 2.2316 in
HE0130$-$4021.\label{fig:fvelplot1}}
\end{figure}

\end{document}